\documentclass[aps,prb,reprint,groupedaddress,superscriptaddress,amsmath,amssymb]{revtex4-2}

\usepackage{graphicx}
\usepackage{blkarray}
\usepackage[colorlinks,urlcolor=blue,linkcolor=blue,anchorcolor=blue,citecolor=blue,bookmarks=false]{hyperref}

\begin{document}

\title{Ligand-SOC enhanced $4f^5$ Kitaev antiferromagnet: Application to $\text{SmI}_3$}

\author{Li-Hao Xia}
\affiliation{National Laboratory of Solid State Microstructures and Department of Physics, Nanjing University, 210093 Nanjing, China}

\author{Yi-Peng Gao}
\affiliation{National Laboratory of Solid State Microstructures and Department of Physics, Nanjing University, 210093 Nanjing, China}

\author{Zhao-Yang Dong}
\email{zhydong@njust.edu.cn}
\affiliation{School of Physics, Nanjing University of Science and Technology, Nanjing 210094, China}

\author{Jian-Xin Li}
\email{jxli@nju.edu.cn}
\affiliation{National Laboratory of Solid State Microstructures and Department of Physics, Nanjing University, 210093 Nanjing, China}
\affiliation{Collaborative Innovation Center of Advanced Microstructures, Nanjing University, 210093 Nanjing, China}

\date{\today}

\begin{abstract} 
The search for Kitaev quantum spin liquids (Kitaev-QSLs) in real materials has mainly focused on $4d$- and $5d$-electron honeycomb systems.  
A recent experimental study on the $4f^5$ honeycomb iodide $\text{SmI}_3$ reported the absence of long-range magnetic order down to $0.1\ \text{K}$, suggesting a possible Kitaev-QSL phase. 
Motivated by the interplay between the complex exchange processes inherent to the $4f^5$ multi-electron configuration and the strong spin-orbit coupling (SOC) of the iodine ligands, we systematically investigate the effective exchange interactions in $\text{SmI}_3$ using the strong coupling expansion method. 
Our findings reveal that bond-dependent SOCs (bond-SOCs), extracted from relativistic density functional theory (DFT) calculations, significantly enhance the antiferromagnetic (AFM) Kitaev interaction, driving the system close to the AFM Kitaev point. A microscopic analysis based on the Slater-Koster approach further indicates that the strong SOC of the iodine ligands (ligand-SOC) is the origin of bond-SOCs and plays a pivotal role in mediating the superexchange processes. 
Additionally, we identify a spin-flop transition induced by the bond-SOCs, where the enhanced AFM Kitaev interactions shift the AFM order from the out-of-plane $[1, 1, 1]$-direction to an in-plane orientation, breaking the $C_3$ rotational symmetry. Linear spin-wave theory (LSWT) further predicts the emergence of gapless modes following the spin-flop transition, indicating enhanced fluctuations and increased instability near the AFM Kitaev point. Our results highlight the crucial role of strong ligand-SOC in stabilizing the dominant AFM Kitaev interactions in $\text{SmI}_3$ and provide valuable insights for discovering new $f$-electron Kitaev-QSL candidates.
\end{abstract}

\maketitle

\section{Introduction}
The concept of quantum spin liquids (QSLs), an exotic and highly entangled quantum phase, has attracted significant interest in condensed matter physics ever since its proposal by Anderson \cite{anderson_resonating_1973,balents_spin_2010,savary_quantum_2016,zhou_quantum_2017,ExperimentalIdentification_Wen_2019,QuantumSpin_Broholm_2020,matsuda_KitaevQuantumSpin_2025}. QSLs are characterized by the absence of long range magnetic order and the presence of strong quantum fluctuations even at zero temperature, which give rise to remarkable collective phenomena such as topological order \cite{wen_topological_1991,QuantumOrders_Wen_2002a}, the fractionalization of spin excitations \cite{sachdev_kagome-_1992}, and emergent anyonic quasiparticles in two-dimensional systems \cite{kitaev_fault-tolerant_2003,kitaev_anyons_2006}. 
A major breakthrough in the study of QSLs came in 2006, when Kitaev proposed a simple yet profound spin model on a honeycomb lattice, providing an exactly solvable QSL ground state \cite{kitaev_anyons_2006}.
The defining feature of this model is its bond-depedent Ising-type spin interactions, known as Kitaev interactions, which introduce considerable spin-exchange frustration, thereby stabilizing a QSL phase. 
However, in real materials, such highly anisotropic Kitaev interactions cannot arise solely from spin degrees of freedom, but require strong entanglement between spin and orbital degrees of freedom.
 
A pioneering work by Jackeli and Khaliullin established an efficient mechanism for realizing dominant Kitaev interactions in $d$-electron spin-orbit-coupled transition-metal compounds, involving the formation of a $J_\text{eff}=1/2$ Kramers doublet ground state and an edge-sharing honeycomb network of ligand octahedra \cite{jackeli_mott_2009}. Based on this framework, numerous $J_\text{eff}=1/2$ $d$-electron Kitaev-QSL candidates have been proposed, including $5d^5$ iridates \cite{KitaevHeisenbergModel_Chaloupka_2010,singh_antiferromagnetic_2010,singh_relevance_2012,ZigzagMagnetic_Chaloupka_2013,modic_realization_2014,FirstPrinciplesStudy_Yamaji_2014,rau_generic_2014,hwan_chun_direct_2015} and $4d^5$ ruthenates \cite{plumb_alpha_RuCl3_2014,kubota_successive_2015,majumder_anisotropic_2015,johnson_monoclinic_2015,banerjee_proximate_2016,wang_theoretical_2017,SpinWaveExcitations_Ran_2017}. More recently, high-spin $S=3/2$ $d^7$ configuration in cobaltates has also been discussed \cite{liu_pseudospin_2018,sano_kitaev-heisenberg_2018,KitaevSpin_Liu_2020,NonKitaevKitaev_Liu_2023}, where the additional spin-active $e_g$ electrons substantially alter the balance between Kitaev and Heisenberg interactions. While there are numerous experimental indications confirming the existence of dominant Kitaev coupling in these materials, most of them are actually magnetically ordered at low enough temperatures, which is attributed to non-Kitaev interactions such as the inherent Heisenberg coupling $J$, symmetry-allowed off-diagonal $\Gamma$ interactions, as well as distortion induced $\Gamma'$ terms \cite{MagneticAnisotropy_Chaloupka_2016,ModelsMaterials_Winter_2017,ConceptRealization_Takagi_2019,HuntingMajorana_Motome_2020,KitaevMaterials_Trebst_2022,ExchangeInteractions_Liu_2022,KitaevPhysics_Rousochatzakis_2024}. The inclusion of these non-Kitaev terms renders the Kitaev-QSL phase narrow and fragile in the phase diagram.
 
Spin-orbit coupling (SOC) plays a fundamental role in generating Kitaev interactions. By entangling spin and orbital degrees of freedom, SOC generates a low-energy doublet that serves as a pseudospin while also modulating exchange energies to stabilize Kitaev interactions. 
An alternative approach to realizing Kitaev physics leverages the strong SOC of heavy ligand anions (ligand-SOC), proposed in high-spin two-dimensional monolayer systems such as $S=1$ $\text{Ni}^{2+}$ and $S=3/2$ $\text{Cr}^{3+}$-based materials \cite{lado_origin_2017,xu_interplay_2018,stavropoulos_microscopic_2019,kim_giant_2019,lee_fundamental_2020,PossibleKitaev_Xu_2020,stavropoulos_magnetic_2021,bandyopadhyay_exchange_2022,riedl_microscopic_2022,OnethirdMagnetization_Shangguan_2023,ChiralSpin_Luo_2024}. In these systems, the ligand-SOC plays a crucial role in generating anisotropic exchange interactions, including the Kitaev type. 
Compared to the atomic SOC of transition-metal ions, the ligand-SOC not only modulates the energies of exchange channels but also introduces additional indirect hopping channels, especially the spin-flip (SF) channels \cite{stavropoulos_magnetic_2021}, that would otherwise be forbidden by spin angular momentum conservation. The effect of these additional hopping channels on Kitaev interactions remains an open question and requires further investigation.

Beyond $d$-electron systems, $f$-electron rare-earth magnets have emerged as promising candidates for realizing Kitaev interactions due to their more localized and highly anisotropic orbitals, as well as stronger SOC \cite{li_kitaev_2017,rau_frustration_2018,MaterialsDesign_Motome_2020,HoneycombRareearth_Luo_2020}. Recent theoretical studies on $4f^1$ rare-earth oxides $\text{A}_2\text{PrO}_3$ ($\text{A}=\text{alkali metals}$) with edge-sharing honeycomb network have demonstrated the presence of dominant Kitaev interactions \cite{jang_antiferromagnetic_2019,jang_computational_2020,jang_exchange_2024}. 
Remarkably, these interactions are antiferromagnetic (AFM), which contrasts sharply with the ferromagnetic (FM) Kitaev interactions typically found in $4d$ and $5d$ candidates.
Under an applied magnetic field, the AFM Kitaev model exhibits a richer phase diagram compared to its FM counterpart. While the FM Kitaev model transitions directly into a high-field polarized phase, the AFM Kitaev model includes an additional intermediate-ﬁeld spin liquid phase between the low-field non-Abelian spin liquid and the high-field polarized phase \cite{banerjee_excitations_2018,janssen_heisenbergkitaev_2019,zhu_robust_2018,gohlke_dynamical_2018,hickey_emergence_2019,zhang_theory_2022}. The diversity and complexity of the AFM Kitaev model’s phase diagram highlights the potential of these $f$-electron rare-earth Kitaev-QSL candidates as an exciting area for further exploration. 

Recent experimental work on the $4f^5$ honeycomb iodide $\text{SmI}_3$ has provided further support for the viability of $f$-electron systems as Kitaev-QSL candidates \cite{ishikawa_SmI3_2022}. The absence of long-range magnetic order down to 0.1 K  suggests that $\text{SmI}_3$ may host a QSL phase. Moreover, the magnetic susceptibility measurements identify a well-defined $\Gamma_7$ Kramers doublet as the $\text{Sm}^{3+}$ ground state, a rare feature still under debate in previous $4f^1$ $\text{Pr}^{4+}$ systems \cite{jang_exchange_2024}. These findings highlight the pressing need for further theoretical and experimental studies to determine the sign and strength of the Kitaev interactions and to elucidate the underlying exchange processes in $\text{SmI}_3$, particularly in the presence of heavy iodine ligands. Given the interplay between complex exchange processes inherent to the $4f^5$ multielectron configuration and the strong ligand-SOC of $\text{I}^-$, $\text{SmI}_3$ serves as an ideal platform to explore their effects on Kitaev interations.  Such investigations could provide deeper insights into the potential of rare-earth magnets for realizing purer and more robust Kitaev interactions.

In this paper, we systematically investigate the $4f^5$ honeycomb iodide $\text{SmI}_3$. Using the strong coupling expansion, we construct a minimal nearest-neighbor (NN) pseudospin model and demonstrate that the strong SOC of the heavy ligand iodine significantly enhances the AFM Kitaev interaction, making it the dominant exchange term. To evaluate the effect of the ligand-SOC, we employ the Slater-Koster approach to derive the effective $f$--$f$ hopping parameters and categorize the hopping channels into two types. Our analysis reveals that the ligand-SOC activates additional hopping channels, introducing bond-dependent SOC effects (bond-SOCs) in the hopping processes that are corroborated by relativistic density functional theory (DFT) calculations. In the strong coupling expansion, the bond-SOCs substantially enhance the AFM Kitaev interaction while slightly suppressing the Heisenberg term. A stronger AFM Kitaev interaction brings the system closer to the Kitaev-QSL phase, where the intrinsic frustration may lead to novel quantum phases.
Furthermore, our investigation uncovers a spin-flop transition induced by the bond-SOCs, wherein the enhanced AFM Kitaev interaction shifts the AFM order from the out-of-plane $[1,1,1]$-direction to an in-plane orientation, breaking the $C_3$ rotational symmetry. Linear spin-wave theory (LSWT) further predicts the emergence of gapless modes upon this spin-flop transition, indicating enhanced fluctuations and increased instability near the AFM Kitaev point. If the AFM order is suppressed by external perturbations such as magnetic fields, possible QSL phases may emerge.    
Our findings underscore the crucial role of the ligand-SOC in mediating the $f$--$p$--$f$ superexchange interactions in $4f$ honeycomb materials with heavy ligands. Moreover, the choice of ligand elements and the effects of orbital hybridizations, such as the $p$--$f$ hybridization, warrant further investigation in the search for $4f$ Kitaev-QSL candidates. This work highlights the importance of ligand-SOC effects and provides valuable insights into the discovery of novel Kitaev-QSL materials.

\section{Model and Method}
\label{sec:model and method}

In this section, we present the theoretical model and methods employed in our study. In Sec.~\ref{subsec:Multiorbital Hamiltonian}, we introduce the full multi-orbital Hamiltonian on the honeycomb lattice of $\text{SmI}_3$. In Sec.~\ref{subsec:Bond structure and ligand-SOC}, we introduce the bond-dependent SOCs and derive the effective $f$--$f$ hopping parameters through a perturbative procedure based on the Slater-Koster approach. In Sec.~\ref{subsec:Strong coupling expansion}, we provide an overview of the strong coupling expansion method and the resulting effective pseudospin Hamiltonian.

\subsection{Multi-orbital Hamiltonian}
\label{subsec:Multiorbital Hamiltonian}
We start from the multi-orbital Hamiltonian of the $4f$ electrons in $\text{Sm}^{3+}$, which is composed of two parts:
\begin{eqnarray}
    H=H_0+H_t,
\label{eq:total Hamiltonian}
\end{eqnarray}
where $H_0$ is the onsite multi-orbital Hamiltonian describing the onsite electron-electron interactions and $H_t$ denotes the kinetic energy terms associated with the electron hopping between nearest-neighbor sites.

The onsite multi-orbital Hamiltonian $H_0$ can be decomposed into three terms:
\begin{eqnarray}
    H_0=H_\text{int}+H_\text{SOC}+H_\text{CEF}.
\label{eq:onsite Hamiltonian}
\end{eqnarray}

The first term, $H_\text{int}$, describes the on-site Coulomb interactions between $f$ electrons and takes the general form:
\begin{eqnarray}
    H_\text{int}=&&\frac{1}{2}\sum_i \sum_{\sigma, \sigma^{\prime}} \sum_{m_{1} m_{2} m_{3} m_{4}} \delta_{m_{1}+m_{2}, m_{3}+m_{4}}\nonumber \\
    && \times\sum_{k=0,2,4,6} F^{k} C^{k}\left(m_{1}, m_{4}\right) C^{k}\left(m_{3}, m_{2}\right)\nonumber\\
    && \times\quad\hat{c}_{i,m_{1} \sigma}^{\dagger} \hat{c}_{i,m_{2} \sigma^{\prime}}^{\dagger} \hat{c}_{i,m_{3} \sigma^{\prime}} \hat{c}_{i,m_{4} \sigma}.
\end{eqnarray}
Here, $F^k$ represents the Slater integral, which accounts for the radial part of the Coulomb interaction, while the angular part $C^k(m,m')$ denotes the Gaunt coeffcients for $k=0,2,4,6$. The relation between the Slater integrals $F^k$ and the more commonly used intraorbital Coulomb interaction $U$ and Hund's coupling $J_h$ in the Hubbard-Kanamori Hamiltonian \cite{georges_strong_2013} is given by $U=F^0$, $J_h=(286F^2+195F^4+250F^6)/6435$ \cite{anisimov_first-principles_1997}.

The second term, $H_\text{SOC}$, describes the spin-orbit coupling of the $4f$ electrons in $\text{Sm}^{3+}$ and is given by:
\begin{eqnarray}
    H_\text{SOC}=\lambda\sum_i L_i \cdot S_i,
\label{eq:onsite SOC Hamiltonian}
\end{eqnarray}
where $L_i$ and $S_i$ denote the total orbital and spin angular momentum operators at site $i$,  respectively, and take the following forms:
\begin{subequations}
\begin{eqnarray}
&&L_i^{(\mu)}=\psi_i^\dagger\cdot(\tau^\mu\otimes\sigma^0)\cdot\psi_i,\\
&&S_i^{(\mu)}=\frac{1}{2}\psi_i^\dagger\cdot(\mathbb{I}_7\otimes\sigma^\mu)\cdot\psi_i.
\end{eqnarray}
\end{subequations}
Here, $\psi_i^\dagger=(\hat{c}_{i,+3\uparrow}^\dagger,\hat{c}_{i,+3\downarrow}^\dagger,\hat{c}_{i,+2\uparrow}^\dagger,\dots,\hat{c}_{i,-3\uparrow}^\dagger,\hat{c}_{i,-3\downarrow}^\dagger)$, where $\hat{c}_{i,m\sigma}^\dagger$ creates an electron with spin $\sigma$ at site $i$ in orbital $m$. The matrices $\tau^\mu$ and $\sigma^\mu$ correspond to the representations of the one-electron orbital operator $l^\mu$ and the Pauli spin matrices ($\mu=x,y,z$), repectively.

For a $4f^5$ configuration under an octahedral crystal field, the crystal field Hamiltonian $H_\text{CEF}$ can be expressed using the operator equivalent method as \cite{stevens_matrix_1952,lea_raising_1962}:
\begin{eqnarray}
    H_\text{CEF}=B_4^0(\hat{O}_4^0+5\hat{O}_4^4),
\label{eq:crystal field Hamiltonian}
\end{eqnarray}
where $\hat{O}_r^s$ (for $s=-r,-r+1,\dots,r$) represents the Stevens multipole operators, and $B_r^s$ are the corresponding coupling constants.

\begin{figure}[!htpb]
\includegraphics[width=0.45\textwidth]{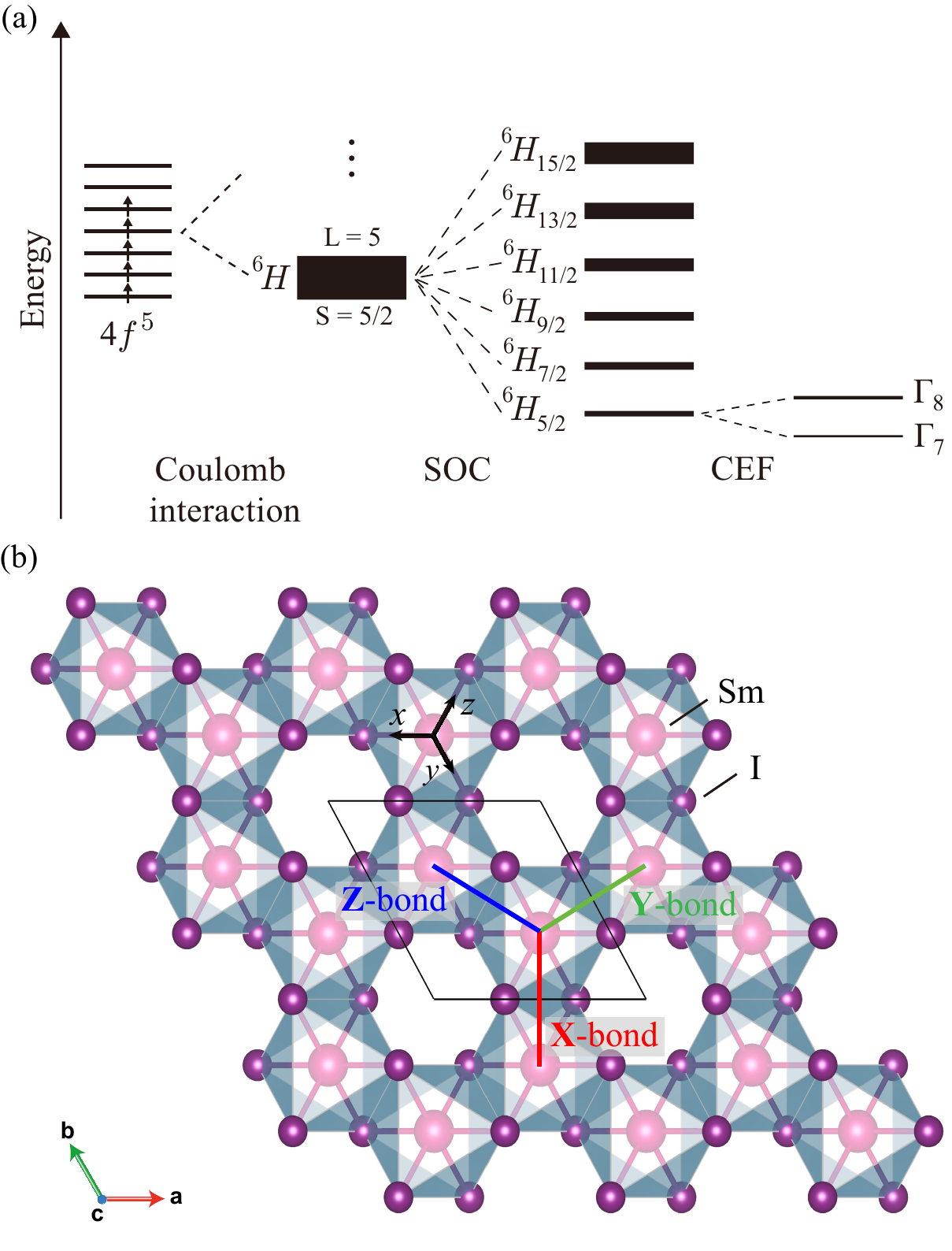}
\caption{ 
(a) Energy levels and multiplet splittings of the $4f^5$-electron configuration, sequentially influenced by Coulomb interaction, spin-orbit coupling, and crystal field effects, ultimately leading to a $\Gamma_7$ Kramers doublet ground state. 
(b) Schematic representation of a $\text{SmI}_3$ honeycomb layer. The red, green, and blue bonds correspond to the $X$-, $Y$- and $Z$-bonds, respectively. Both the crystallographic ($abc$) coordinate system and the local cubic ($xyz$) coordinate system are deﬁned.}
\label{fig:SmI3_structure}
\end{figure}

The kinetic energy term $H_t$, which is treated perturbatively in our work, is given by: 
\begin{eqnarray}
H_t=\sum_{<i,j>}\psi_i^\dagger\cdot\mathcal{T}_{ij}^{(\Lambda)}\cdot\psi_j+\text{H.c.},
\end{eqnarray}
where $\psi_i^\dagger=(\hat{c}_{i,\xi\uparrow}^\dagger,\hat{c}_{i,\xi\downarrow}^\dagger,\hat{c}_{i,\eta\uparrow}^\dagger,\hat{c}_{i,\eta\downarrow}^\dagger,\dots,\hat{c}_{i,\gamma\uparrow}^\dagger,\hat{c}_{i,\gamma\downarrow}^\dagger)$ represents the electron creation operators with $\xi,\eta,\zeta,A,\alpha,\beta$ and $\gamma$ denoting the seven types of $4f$ orbitals in the cubic harmonic basis \cite{takegahara_slater-koster_1980}. The matrix $\mathcal{T}_{ij}^{(\Lambda)}$ encodes the hopping parameters between nearest-neighbor sites $i$ and $j$ along the $\Lambda$-bond ($\Lambda=X,Y,Z$, as shown in Fig.\ref{fig:SmI3_structure}~(b)).

\subsection{Bond structure and ligand-SOC}
\label{subsec:Bond structure and ligand-SOC}
 
By selectively enabling and disabling SOCs in our DFT calculations, we obtain two distinct hopping matrices:  $\mathcal{T}_{ij}^{\text{SOCs},(\Lambda)}$, calculated with SOCs, and $\mathcal{T}_{ij}^{(\Lambda)}$, calculated without SOCs (for details, see Appendix \ref{appendix:ab initio results}). The difference between them can be interpreted as the inclusion of a type of bond-dependent SOC arising in the hopping processes:
\begin{eqnarray}
\mathcal{T}_{ij}^{\text{SOCs},(\Lambda)}=\mathcal{T}_{ij}^{(\Lambda)}+\mathcal{T}_{ij}^{\text{bond-SOCs},(\Lambda)}.
\end{eqnarray}
The bond-SOC hopping parameters $\mathcal{T}_{ij}^{\text{bond-SOCs},(\Lambda)}$, which consists of a few non-negligible spin-dependent terms, can be further parameterized by introducing the vector matrix $\vec{\lambda}$ as follows \cite{riedl_microscopic_2022}:
\begin{eqnarray}
\label{eq:bond-SOCs}
\mathcal{T}_{ij}^{\text{bond-SOCs},(\Lambda)}=\frac{1}{2}\vec{\lambda}^{(\Lambda)}\cdot\vec{\sigma},
\end{eqnarray}
which resembles the general form of the atomic SOC.

The non-relativistic hopping parameters $\mathcal{T}_{ij}^{(\Lambda)}$ and the bond-SOC coupling matrix $\vec{\lambda}$ on a $Z$-bond are listed in Table \ref{tab:hopping parameters} and Table \ref{tab:relativistic hopping parameters} in Appendix \ref{appendix:hopping parameters}.

To elucidate the microscopic origin of bond-SOCs, we analytically derive the hopping matrix $\mathcal{T}_{ij}^{\text{SOCs},(\Lambda)}$ using the Slater-Koster approach. Given the highly localized nature of the $4f$ orbitals and the complexity of introducing additional Slater-Koster integrals, we neglect direct $f$--$f$ hoppings and instead focus on indirect $f$--$p$--$f$ hoppings mediated by the $5p$ orbitals of $\text{I}^-$ ligands. Assuming an ideal octahedral coordination environment, the $f$--$p$ hopping matrix $T_{fp}$ can be expressed in terms of Slater-Koster parameters. For two Sm--I sites aligned along the vector $(-1,0,0)$, the hopping matrix takes the following form:
\begin{equation}
\label{eq:f_to_p_hopping_matrix}
    T_{fp}^{(-1,0,0)}= \ \ 
\begin{blockarray}{ccccccc}
p_x & p_y & p_z\\
\begin{block}{(ccc)cccc}
0& 0& 0& &&f_\xi\\
0& -t_0& 0& &&f_\eta\\
0& 0& t_0& &&f_\zeta\\
0& 0& 0& &&f_A\\
t_2& 0& 0& &&f_\alpha\\
0& -t_1& 0& &&f_\beta\\
0& 0& -t_1& &&f_\gamma\\ 
\end{block}    
\end{blockarray},
\end{equation}
\vspace{12pt}
where $t_0=\frac{1}{2}\sqrt{\frac{5}{2}}t_{pf\pi}$, $t_1=\frac{1}{2}\sqrt{\frac{3}{2}}t_{pf\pi}$, $t_2=t_{pf\sigma}\ (t_{pf\pi}<0,t_{pf\sigma}>0)$.

To derive the effective $f$--$f$ hopping parameters $\mathcal{T}_{ij}^{\text{eff},(\Lambda)}$ on a $\Lambda$-bond, we integrate out the $5p$ orbitals of the $\text{I}^{-}$ ligands through a perturbative procedure, incorporating the effect of strong ligand-SOCs as demonstrated in the case of $\text{CrI}_3$ \cite{stavropoulos_magnetic_2021}, and obtain:
\begin{eqnarray}
\mathcal{T}_{ij}^{\text{eff},(\Lambda)}=\sum_k\sum_a\frac{T_{ik}|a\rangle\langle a|T_{kj}}{\Delta E_a}+\text{h.c.}
,
\end{eqnarray}

where the summation runs over all single-hole intermediate states $a$ of the two ligand sites $k$. The energy difference is given by $\Delta E_a = \Delta+\lambda_p $ or $\Delta-\lambda_p/2$, where $\Delta$ represents the atomic energy difference between the $4f$ and $5p$ orbitals, and $\lambda_p$ denotes the atomic SOC of the ligand $5p$ orbitals. This approach is validated by our DFT calculations, which show minimal hybridization between the $4f$ orbitals of $\text{Sm}^{3+}$ and the $5p$ orbitals of $\text{I}^-$ (see Appendix \ref{appendix:ab initio results} for details). After simplification, the effective $f$--$f$ hopping matrix on a $Z$-bond takes the following form:

\begin{widetext}
\begin{equation}
\label{eq:hopping_parameters_with_ligand_SOC}
\mathcal{T}_{ij}^{\text{eff},(Z)}=t_\text{eff}
\left(
\begin{array}{ccccccc}
0_{2\times2} & i\frac{r}{2}\sigma^z & i\frac{r}{2}\sigma^y & 0_{2\times2} &\frac{t_2}{t_0}\sigma^0 & i\frac{r}{2}\frac{t_1}{t_0}\sigma^z & -i\frac{r}{2}\frac{t_1}{t_0}\sigma^y  \\
-i\frac{r}{2}\sigma^z & 0_{2\times2} & -i\frac{r}{2}\sigma^x & 0_{2\times2} & i\frac{r}{2}\frac{t_1}{t_0}\sigma^z & -\frac{t_2}{t_0}\sigma^0 & -i\frac{r}{2}\frac{t_1}{t_0}\sigma^x \\
-i\frac{r}{2}\sigma^y & i\frac{r}{2}\sigma^x & -2\sigma^0 & 0_{2\times2} & i\frac{r}{2}\frac{t_1+t_2}{t_0}\sigma^y & i\frac{r}{2}\frac{t_1+t_2}{t_0}\sigma^x  & 0_{2\times2} \\
0_{2\times2} & 0_{2\times2} & 0_{2\times2} & 0_{2\times2} & 0_{2\times2} & 0_{2\times2} & 0_{2\times2} \\
\frac{t_2}{t_0}\sigma^0 & -i\frac{r}{2}\frac{t_1}{t_0}\sigma^z & -i\frac{r}{2}\frac{t_1+t_2}{t_0}\sigma^y & 0_{2\times2} &-2\frac{t_1t_2}{t_0^2}\sigma^0 & -i\frac{r}{2}\frac{t_1^2+t_2^2}{t_0^2}\sigma^z & i\frac{r}{2}\frac{t_1(t_1-t_2)}{t_0^2}\sigma^y  \\
-i\frac{r}{2}\frac{t_1}{t_0}\sigma^z & -\frac{t_2}{t_0}\sigma^0 & -i\frac{r}{2}\frac{t_1+t_2}{t_0}\sigma^x & 0_{2\times2} & i\frac{r}{2}\frac{t_1^2+t_2^2}{t_0^2}\sigma^z & -2\frac{t_1t_2}{t_0^2}\sigma^0 & -i\frac{r}{2}\frac{t_1(t_1-t_2)}{t_0^2}\sigma^x \\
i\frac{r}{2}\frac{t_1}{t_0}\sigma^y & i\frac{r}{2}\frac{t_1}{t_0}\sigma^x & 0_{2\times2} & 0_{2\times2} & -i\frac{r}{2}\frac{t_1(t_1-t_2)}{t_0^2}\sigma^y & i\frac{r}{2}\frac{t_1(t_1-t_2)}{t_0^2}\sigma^x & 2\frac{t_1^2}{t_0^2}\sigma^0 \\
\end{array}
\right)
,
\end{equation}
\end{widetext}
where the basis is $\lbrace f_\xi,f_\eta,f_\zeta,f_A,f_\alpha,f_\beta,f_\gamma\rbrace\otimes\lbrace\uparrow,\downarrow\rbrace$. $t_\text{eff}=\frac{t_0^2}{3}(\frac{2}{\Delta-\lambda_p/2}+\frac{1}{\Delta+\lambda_p})$ is the effective $f$--$p$ hopping integral, and $r=\frac{2\lambda_p}{2\Delta+\lambda_p}$ quantifies the ratio of the ligand-SOC $\lambda_p$ to the atomic energy difference $\Delta$. When $\lambda_p=0$, Eq.~(\ref{eq:hopping_parameters_with_ligand_SOC}) reduces to:
\begin{equation}
\label{eq:hopping_parameters_without_ligand_SOC}
\mathcal{T}_{ij}^{\text{eff},(Z)}=
\left(
\begin{array}{ccccccc}
0 &0 &0 &0 &\frac{t_0t_2}{\Delta} &0 &0 \\
0 &0 &0 &0 &0 &-\frac{t_0t_2}{\Delta} &0 \\
0 &0 &-\frac{2t_0^2}{\Delta} &0 &0 &0 &0 \\
0 &0 &0 &0 &0 &0 &0 \\
\frac{t_0t_2}{\Delta} &0 &0 &0 &-\frac{2t_1t_2}{\Delta} &0 &0 \\
0 &-\frac{t_0t_2}{\Delta} &0 &0 &0 &-\frac{2t_1t_2}{\Delta} &0 \\
0 &0 &0 &0 &0 &0 &\frac{2t_1^2}{\Delta} \\
\end{array}
\right)\otimes\sigma^0
.
\end{equation}
By comparing Eqs.~(\ref{eq:hopping_parameters_with_ligand_SOC}) and (\ref{eq:hopping_parameters_without_ligand_SOC}) with the hopping parameters extracted from our relativistic and non-relativistic DFT calculations, we find good agreement in the main terms, except for two non-negligible $\sigma^0$ components in the $\xi\xi$ and $\eta\eta$ hopping channels, which will be addressed in Sec.~\ref{sec:discussion and summary}. This suggests that the bond-SOCs primarily originate from the strong ligand-SOC of $\text{I}^-$. We set the Slater-Koster integrals as $t_{pf\sigma}=0.24\ \text{eV}$, $t_{pf\pi}/t_{pf\sigma}=-0.5$ with $\lambda_p=0.63\ \text{eV}$ and $\Delta=0.8\ \text{eV}$, to fit the DFT results. The origin of the additional hopping terms that appear when the ligand-SOC is included will be discussed later in Sec.~\ref{subsec:effect of ligand-SOC}.

\subsection{Strong coupling expansion}
\label{subsec:Strong coupling expansion}

We now discuss how to project the multi-orbital Hamiltonian onto the lowest energy subspace to obtain the minimal effective pseudospin model.

According to the high-temperature magnetic susceptibility measurements of $\text{SmI}_3$, the $\text{Sm}^{3+}$ ion exhibits a well-defined $\Gamma_7$ Kramers doublet as its ground states \cite{ishikawa_SmI3_2022}. This indicates that the Russell-Saunders coupling scheme provides an appropriate framework for describing the energy-level diagram, as shown in Fig.~\ref{fig:SmI3_structure}~(a). The $4f^5$ multi-electron manifold of $\text{Sm}^{3+}$ is first split by the Coulomb interaction into different multiplets labeled by term symbols $^{2S+1}L$, where $L$ and $S$ denote the total orbital and spin angular momentum quantum numbers, respectively. The lowest 66-fold degenerate $^6H$ manifold ($L=5$, $S=5/2$) is then split by the atomic SOC into $^6H_j$ manifolds, where $j$ denotes the total angular momentum ($j=\frac{5}{2},\frac{7}{2},\dots,\frac{15}{2}$). The octahedral crystal field further splits the $^6H_{5/2}$ sextet into a $\Gamma_8$ quartet and the lowest $\Gamma_7$ Kramers doublet, which is characterized by an effective angular momentum $j_\text{eff}=1/2$. The wave functions of the $\Gamma_7$ Kramers doublet in the $|j,j^z\rangle$ basis are expressed as:
\begin{eqnarray}
    &&|\tilde{\pm}\rangle =|j_{\text{eff}}^z=\pm\frac{1}{2}\rangle \nonumber\\ &&=\frac{i}{\sqrt{6}}|j =\frac{5}{2},j^z=\pm\frac{5}{2}\rangle-i\sqrt{\frac{5}{6}}|j=\frac{5}{2},j^z=\mp\frac{3}{2}\rangle \nonumber .\\
\label{eq:Kramers doublet}
\end{eqnarray}
The total angular momentum operator $j^\mu$ behaves like a spin-$\frac{1}{2}$ operator within this doublet subspace, differing only in the coefficient. Accordingly, we define the pseudospin operator $\tilde{S}^\mu$ as:
\begin{equation}
    \tilde{S}^\mu=-\frac{3}{5}j^\mu=\frac{1}{2}\sigma^\mu
    ,
\label{eq:definition of the pseudospin}
\end{equation} 
% where $\sigma^\mu$ represents the $\mu$ component of the Pauli matrices.
 
In the large-$U$ strong-coupling limit, treating the hopping term $H_t$ in Eq.~(\ref{eq:total Hamiltonian}) as perturbation, we project out the high-energy states through the second-order perturbation approximation. The resulting effective $\tilde{S}=1/2$ spin-spin interaction Hamiltonian on a $\Lambda$-bond ($\Lambda=X,Y,Z$) is given by:
\begin{equation}
    H_{ij}^{\text{eff},(\Lambda)}=\sum_{\alpha,\beta}\sum_{n}|\alpha\rangle\frac{\langle\alpha|H_t^{(\Lambda)}|n\rangle\langle n|H_t^{(\Lambda)}|\beta\rangle}{E_0-E_n}\langle\beta|
    ,
\label{eq:effective Hamiltonian}
\end{equation}
where $|\alpha\rangle,|\beta\rangle \in\lbrace|\tilde{+}\rangle_i,|\tilde{-}\rangle_i\rbrace\otimes\lbrace|\tilde{+}\rangle_j,|\tilde{-}\rangle_j\rbrace$. Here, $|n\rangle$ denotes one of the intermediate states $|4f^4 \rangle_i \otimes |4f^6 \rangle_j$ or $|4f^6 \rangle_i \otimes |4f^4 \rangle_j$, while $E_0$ and $E_n$ represent the energy of the ground state and the intermediate states, respectively. The primary challenge in this perturbative projection arises from the large dimension of the intermediate-state Hilbert space, which can reach up to $2C_{14}^4C_{14}^6=6,012,006$. This is several orders of magnitude larger than that in the prior $d$-electron cases, rendering analytic analysis nearly impossible. Therefore, we employ numerical calculations to obtain the effective Hamiltonian and resort it in our analysis. 
The $4\times4$ effective Hamiltonian in Eq.~(\ref{eq:effective Hamiltonian}) can be reorganized into a generalized microscopic spin Hamiltonian of the form: $H_{ij}^\text{eff}=\sum_{ij,\alpha\beta}J_{ij}^{\alpha\beta}\tilde{S}_i^{\alpha}\tilde{S}_j^{\beta}$, where $J_{ij}^{\alpha\beta}$ represents the exchange coupling coefficients. Considering the symmetry-allowed terms on a $\Lambda$-bond between the nearest-neighbor (NN) sites $i$ and $j$, the effective pseudospin Hamiltonian can be reduced to the form of $J-K-\Gamma-\Gamma'$ model (for example, for the $Z$-bond): 
\begin{equation}
    H_{<ij>}^{\text{eff},(Z)}=\mathbf{\tilde{S}}^{\mathrm{T}}\cdot \left(
    \begin{array}{ccc}
    J &\Gamma &\Gamma' \\
    \Gamma &J &\Gamma' \\
    \Gamma' &\Gamma' &J+K \\
    \end{array}
    \right)\cdot\mathbf{\tilde{S}} ,
\end{equation}
where $\mathbf{\tilde{S}}^{\mathrm{T}}=(\tilde{S}^x,\tilde{S}^y,\tilde{S}^z)$, and the interaction terms include the bond-isotropic Heisenberg exchange $J$, the anisotropic Kitaev exchange $K$, as well as the off-diagonal exchange terms $\Gamma$ and $\Gamma'$. Considering the $C_3$ rotational symmetry of the honeycomb lattice (see Fig.~\ref{fig:SmI3_structure}~(b)),  $H^{\text{eff},{(X)}}$ and $H^{\text{eff},{(Y)}}$ can be obtained by the cyclic permutation of $\lbrace xyz \rbrace$ in $H^{\text{eff},{(Z)}}$.

\section{results}
\label{sec:results}

In this section, we present the results of our study on the honeycomb compound $\text{SmI}_3$. In Sec.~\ref{subsec:Exchange couplings}, we calculate the effective exchange couplings as functions of the Coulomb interaction $U$ and Hund's coupling $J_h$, both with and without the bond-SOCs, and further decompose these couplings into contributions from various exchange channels. In Sec.~\ref{subsec:effect of ligand-SOC}, we provide a microscopic analysis of the ligand-SOC effect, categorizing the hopping channels into two distinct types and analyzing their individual and combined contributions. Finally, in Sec.~\ref{subsec:Spin excitation spectra}, we employ linear spin-wave theory (LSWT) to investigate the AFM order and spin excitation spectra, offering understandings into the magnetic properties of $\text{SmI}_3$.

\begin{figure*}[!htbp]
\includegraphics[width=\textwidth]{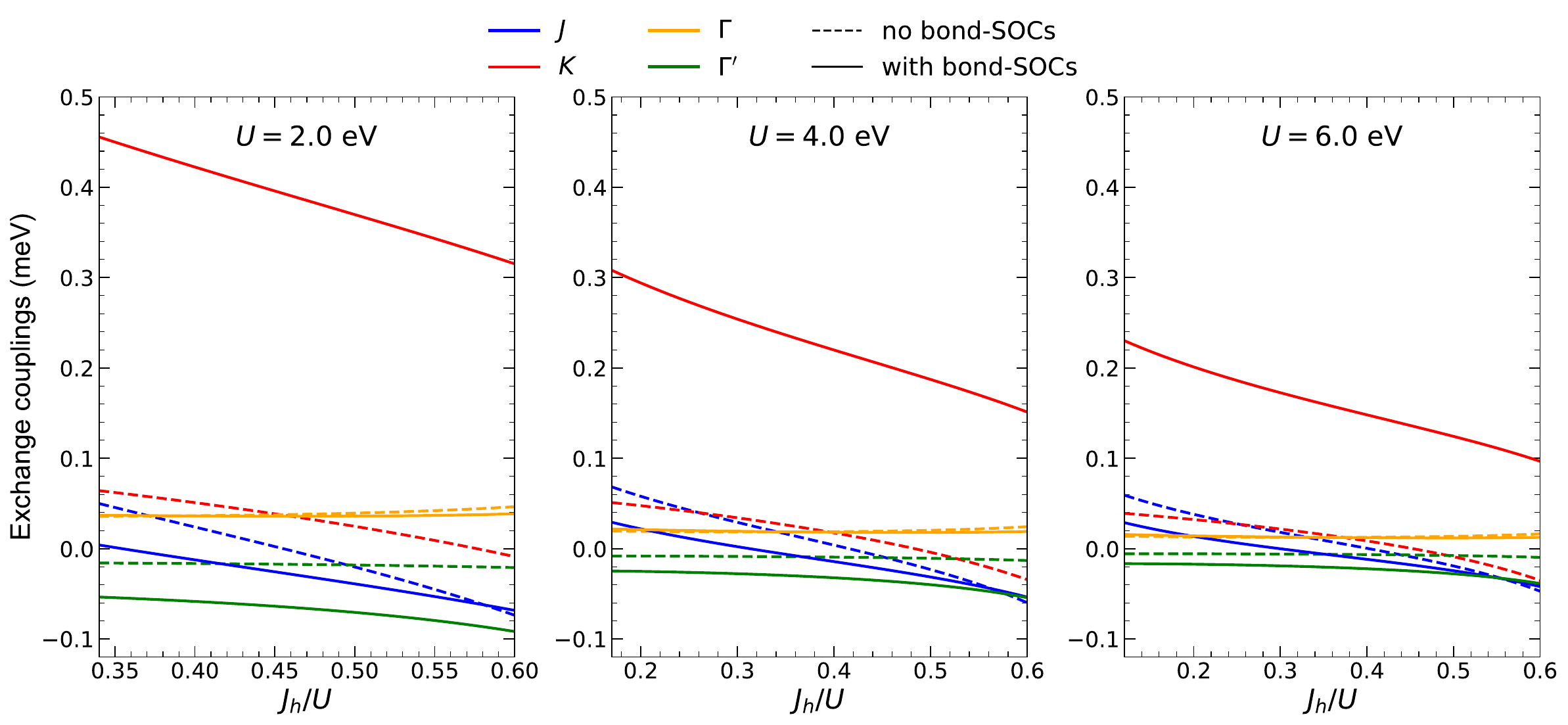}
\caption{ 
Exchange couplings of the effective pseudospin Hamiltonian on a $Z$-bond as a function of the ratio $J_h/U$ for $U=2,4,6 \  \text{eV}$. Solid and dashed lines correspond to calculations using hopping parameters with and without the bond-SOCs, respectively. The atomic SOC strength of $\text{Sm}^{3+}$ and the crystal field parameter are set to $\lambda=36.9\ \text{meV}$ and $B_4^0=4.8\times10^{-2}\ \text{meV}$, respectively, to match the experimentally observed energy splittings. Results for small $J_h/U$, which deviate significantly from the Russell-Saunders coupling scheme, are omitted for clarity.}
\label{fig:Exchange couplings}
\end{figure*}

\subsection{Exchange couplings}
\label{subsec:Exchange couplings}

\begin{figure*}[!htbp]
\includegraphics[width=\textwidth]{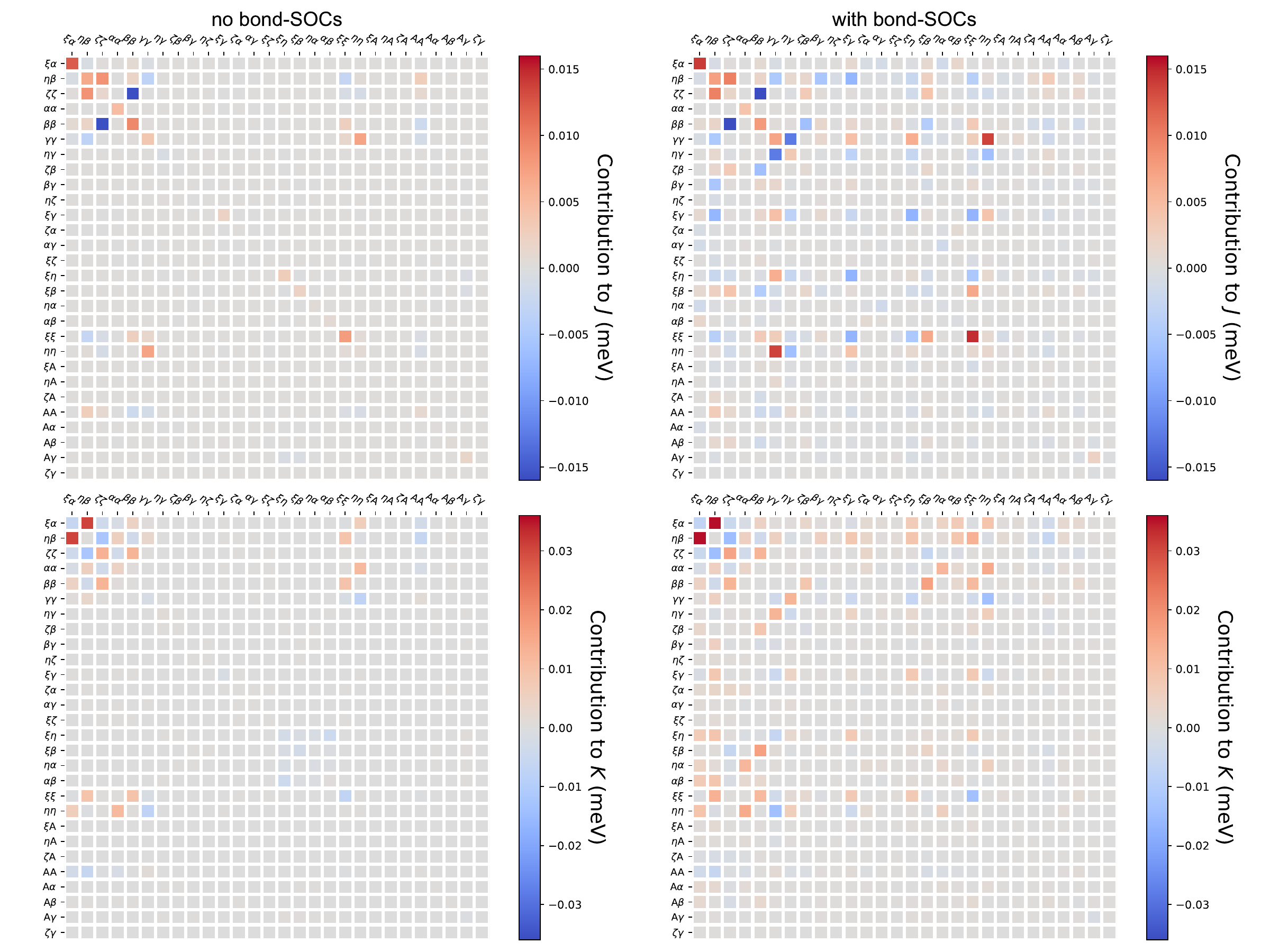}
\caption{
Heatmaps illustrating the contributions from various exchange channels to the Heisenberg coupling $J$ and the Kitaev coupling $K$ on a $Z$-bond, with and without bond-SOCs. The label pair $(uv;u'v')$ indicates the exchange channel composed of hopping parameters $t_{iu\sigma,jv\sigma'}$ and $t_{iu'\sigma,jv'\sigma'}$ on a $Z$-bond, where $u,v,u',v'\in \{\xi,\eta,\zeta,A,\alpha,\beta,\gamma$\} and $\sigma,\sigma' \in \{ \uparrow,\downarrow \}$. Red units indicate AFM contributions, while blue units correspond to FM contributions. The onsite Coulomb energy  is set to $U=5.53$ eV, with a Hund’s coupling ratio of $J_h/U=0.15$. The atomic SOC strength of $\text{Sm}^{3+}$ and the crystal field parameter are set to $\lambda=36.9\ \text{meV}$ and $B_4^0=4.8\times10^{-2}\ \text{meV}$, respectively.}
\label{fig:reorder_contributions_J_K}
\end{figure*}

We begin our analysis by performing the strong coupling expansion using the nearest-neighbor hopping parameters extracted from our DFT calculations. The resulting effective exchange coupling constants for different onsite Coulomb interactions, $U=2,4,6\ \text{eV}$, as functions of the ratio of the Hund’s coupling $J_h$ to $U$ are shown in Fig.~\ref{fig:Exchange couplings}. The atomic SOC strength $\lambda$ of $\text{Sm}^{3+}$ in Eq.~(\ref{eq:onsite SOC Hamiltonian}) and the crystal field parameter $B_4^0$ in Eq.~(\ref{eq:crystal field Hamiltonian}) are determined by matching the calculated energy splittings with experimental results \cite{ishikawa_SmI3_2022}, yielding $\lambda=36.9\ \text{meV}$ and $B_4^0=0.048\ \text{meV}$. Notably, the wave functions of the Kramers doublet in Eq.~(\ref{eq:Kramers doublet}) and the $|4f^4\rangle$, $|4f^6\rangle$ intermediate states in Eq.~(\ref{eq:effective Hamiltonian}) are obtained by numerically diagonalizing the onsite Hamiltonian in Eq.~(\ref{eq:onsite Hamiltonian}) for various multi-electron configurations. 
Our calculations demonstrate that at small values of $J_h/U$, the energy levels of the $4f^5$ multiplets undergo a reordering, leading the ground state to deviate from the well-defined $\Gamma_7$ manifold described by the Russell-Saunders coupling scheme and shift toward the intermediate coupling regime. To preserve the consistency of the pseudospin definition in Eq.~(\ref{eq:definition of the pseudospin}), we explicitly exclude these parameter ranges from subsequent analysis.

A comparison of the effective exchange couplings with (solid lines) and without (dashed lines) the bond-SOCs in Fig.~\ref{fig:Exchange couplings} reveals the significant impact of the bond-SOCs, particularly on the Kitaev interaction $K$. In the absence of the bond-SOCs, $K$ is comparable in magnitude to the Heisenberg interaction $J$ and the symmetric off-diagonal interaction $\Gamma$, all of which remain below $0.1\ \text{meV}$. In this case, $K$ transitions from AFM to FM as $J_h/U$ increases. However, when the bond-SOCs are included, $K$ is significantly enhanced compared to the other interactions and remains AFM across all the parameter regions considered. This result suggests that the bond-SOCs play a crucial role in stabilizing a dominant AFM Kitaev interaction.

Based on experimental spectroscopic measurements, we set the intraorbital Coulomb interaction to $U=5.53\ \text{eV}$ \cite{cox_study_1981} and the Hund's coupling to $J_h=0.837\ \text{eV}$ \cite{wybourne_spectroscopic_1965}, corresponding to a ratio of $J_h/U=0.15$. The calculated exchange coupling constants for this parameter set, with and without the bond-SOCs, are listed in Table~\ref{tab:exchange parameters}. A comparison with the $J-K-\Gamma-\Gamma'$ phase diagrams obtained via exact diagonalization \cite{rusnaifmmode_checkcelse_cfiko_kitaev-like_2019} reveals that, in the absence of the bond-SOCs, the coupling constants fall within the AFM region. In contrast, when the bond-SOCs are included, the coupling constants shift closer to the boundary between the AFM and vortex regions. More importantly, they move significantly toward the AFM Kitaev point, highlighting the strong potential for realizing an AFM Kitaev-QSL phase.

\begin{table}
\caption{
Exchange coupling constants calculated for $U=5.53\ \text{eV}$ and $J_h/U=0.15$ on a $Z$-bond with and without the bond-SOCs. The atomic SOC strength of $\text{Sm}^{3+}$ and the crystal field parameter are set to $\lambda=36.9\ \text{meV}$ and $B_4^0=4.8\times10^{-2}\ \text{meV}$, respectively. The unit is $\text{meV}$.}
\label{tab:exchange parameters}
\begin{ruledtabular}
\begin{tabular}{ccccccc}
&&$J$ &$K$ &$\Gamma$ &$\Gamma'$ \\ \hline
&with bond-SOCs &0.025 &0.227 &0.017 &-0.016  \\ \hline
&without bond-SOCs &0.054 &0.038 &0.015 &-0.006 \\
\end{tabular}
\end{ruledtabular}
\end{table}

\begin{figure*}[!htbp]
\includegraphics[width=\textwidth]{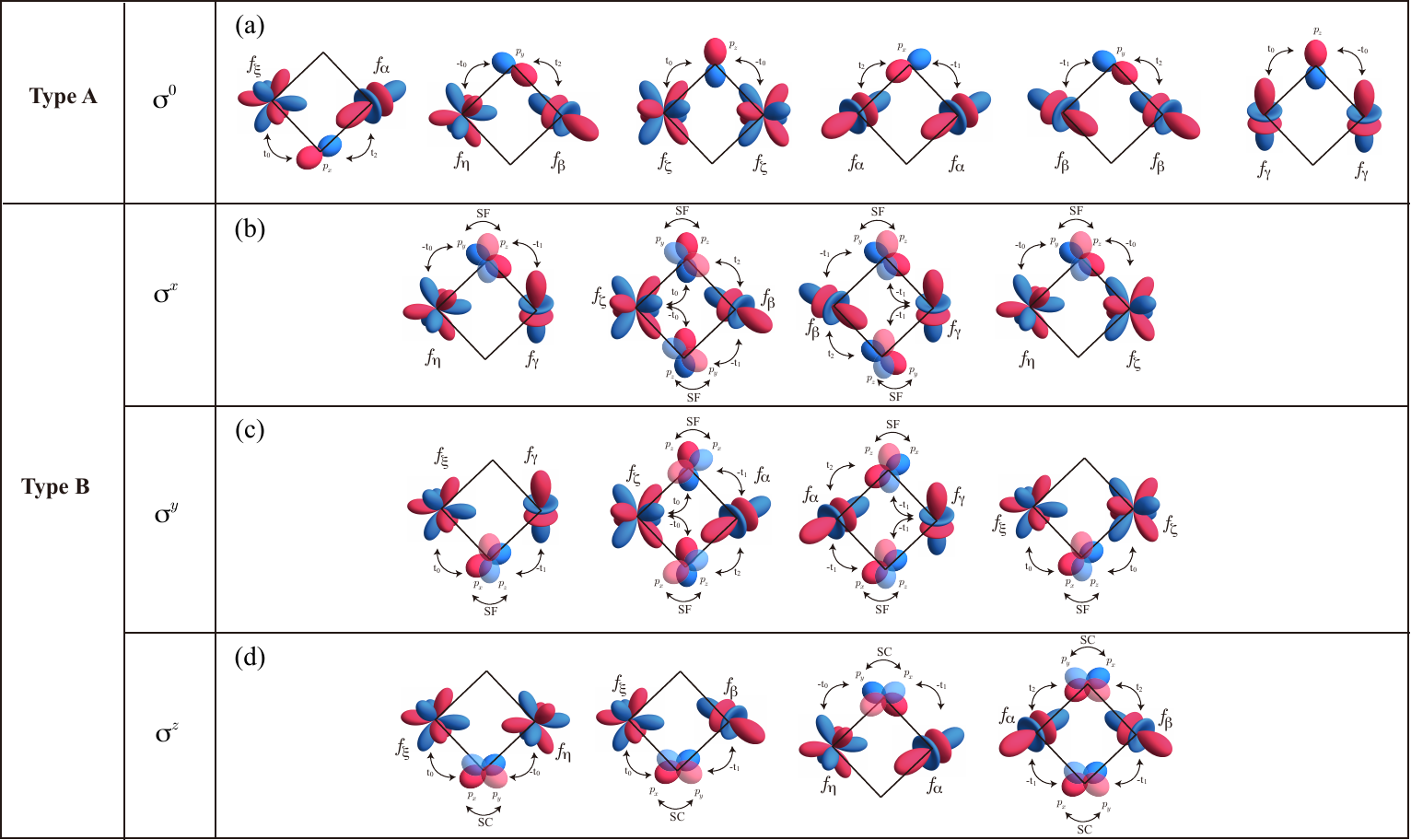}
\caption{
Schematic representation of the indirect hopping channels on a $Z$-bond, corresponding to the nonzero terms in Eq.~(\ref{eq:hopping_parameters_with_ligand_SOC}) categorized by the Pauli matrix components of their hopping parameters. (a) $\sigma^0$ channels: $\xi\alpha$, $\eta\beta$, $\zeta\zeta$, $\alpha\alpha$, $\beta\beta$, $\gamma\gamma$ (b) $\sigma^x$ channels: $\eta\gamma$, $\zeta\beta$, $\beta\gamma$, $\eta\zeta$. (c) $\sigma^y$ channels: $\xi\gamma$, $\zeta\alpha$, $\alpha\gamma$, $\xi\zeta$. (d) $\sigma^z$ channels: $\xi\eta$, $\xi\beta$, $\eta\alpha$, $\alpha\beta$. These channels can be further categorized into spin-flip (SF) and spin-conserving (SC) channels, depending on whether a spin flip occurs at the ligand sites.}
\label{fig:hopping_channels}
\end{figure*}
 
To gain deeper insight into the exchange processes, we decompose the exchange couplings into contributions from different exchange channels. Heatmaps illustrating these contributions to the Heisenberg coupling $J$ and the Kitaev coupling $K$ on a $Z$-bond are shown in Fig.~\ref{fig:reorder_contributions_J_K}. Each unit labeled as $(uv;u'v')$ represents an exchange channel in which a $4f$ electron hops from orbital $u$ on site $i$ to orbital $v$ on site $j$, then returns from orbital $v'$ on site $j$ to orbital $u'$ on site $i$, including the corresponding mirror processes. In the absence of the bond-SOCs, the dominant FM contribution to $J$ arises from the $(\beta\beta;\zeta\zeta)$ channel, while AFM contributions primarily originate from channels such as $(\xi\alpha;\xi\alpha)$, 
$(\eta\beta;\eta\beta)$, $(\eta\beta;\zeta\zeta)$, $(\beta\beta;\beta\beta)$, $(\gamma\gamma;\eta\eta)$, and $(\xi\xi;\xi\xi)$. Meanwhile, the AFM Kitaev interaction $K$ is predominantly contributed by the $(\xi\alpha;\eta\beta)$ channel, similar to the $4f^1$ case discussed in \cite{jang_antiferromagnetic_2019,jang_computational_2020}. 
 
When the bond-SOCs are included, numerous additional exchange channels emerge, substantially modifying the exchange interactions. The microscopic origin of these new channels will be discussed in detail in Sec.~\ref{subsec:effect of ligand-SOC}.
For the Heisenberg coupling $J$, several new exchange channels---such as  $(\gamma\gamma;\eta\gamma)$, $(\eta\beta;\xi\gamma)$, $(\xi\gamma;\xi\eta)$, and $(\xi\gamma;\xi\xi)$---introduce non-negligible FM contributions, which suppress the previously dominant AFM components. In contrast, nearly all the additional channels contribute to the AFM Kitaev coupling $K$, further enhancing the AFM Kitaev interaction and establishing it as the dominant term. These findings highlight the crucial role of the bond-SOCs in stabilizing a dominant AFM Kitaev model---an effect previously unexplored in $f$-electron systems.

\subsection{Ligand-SOC effect}
\label{subsec:effect of ligand-SOC}

Now, we analyze the microscopic origin of the bond-SOCs in $\text{SmI}_3$, which we attribute to the strong ligand-SOC arising from the heavy iodine ligands.  Similar effects have been extensively studied in $3d$ layered transition-metal compounds with heavy ligands, such as $\text{NiI}_2$ \cite{stavropoulos_microscopic_2019,riedl_microscopic_2022} and $\text{CrI}_3$ \cite{lado_origin_2017,xu_interplay_2018,stavropoulos_magnetic_2021}, where the ligand-SOC plays a crucial role in superexchange processes and the formation of anisotropic exchange interactions. The SOC Hamiltonian for the $p$-orbital electrons in the ligand ions is given by:
\begin{equation}
\label{eq:p orbotal SOC}
    H_{\text{SOC}}^{p}= \ \ 
\begin{blockarray}{cccccccccc}
p_{x,\uparrow} & p_{x,\downarrow} & p_{y,\uparrow} &p_{y,\downarrow} &p_{z,\uparrow} &p_{z,\downarrow}\\
\begin{block}{(cccccc)cccc}
0& 0& \frac{i\lambda_p}{2} & 0& 0& -\frac{\lambda_p}{2}& &&p_{x,\uparrow} \\
0& 0& 0& -\frac{i\lambda_p}{2}& \frac{\lambda_p}{2}& 0& &&p_{x,\downarrow} \\
-\frac{i\lambda_p}{2}& 0& 0& 0& 0& \frac{i\lambda_p}{2}& &&p_{y,\uparrow} \\
0& \frac{i\lambda_p}{2}& 0& 0& \frac{i\lambda_p}{2}& 0& &&p_{y,\downarrow} \\
0& \frac{\lambda_p}{2}& 0& -\frac{i\lambda_p}{2}& 0& 0& &&p_{z,\uparrow} \\
-\frac{\lambda_p}{2}& 0& -\frac{i\lambda_p}{2}& 0& 0& 0& &&p_{z,\downarrow} \\
\end{block}    
\end{blockarray}.
\end{equation}
The ligand-SOC splits the initially degenerate $p$ orbitals and entangles them with the spin degree of freedom, leading to two key effects: (i) modulation of the eigenenergies of the spin-orbit-entangled states, and (ii) the emergence of additional off-diagonal spin-dependent terms that enable spin-flip (SF) processes within the ligand ions. Both effects influence the $f$--$p$--$f$ superexchange processes. Notably, the SF terms have been proposed as the primary origin of the Kitaev interactions in $\text{CrI}_3$ \cite{stavropoulos_magnetic_2021}. 
Compared to the $d$-electron Kitaev-QSL candidates with heavy ligands, the $4f^5$ system $\text{SmI}_3$ exhibits additional complexities. First, the wave functions of both the $4f^5$ Kramers doublet and the $4f^{4}/ 4f^{6}$ intermediate states are more intricate in the electron-occupied basis, containing multiple multi-electron configurations where $f$ orbitals and spins are strongly entangled. Second, the $f$ orbitals exhibit higher spatial anisotropy, which could introduce additional degrees of freedom in the $f$--$p$--$f$ superexchange processes. The interplay between these two factors renders the exchange processes in $\text{SmI}_3$ particularly complex and challenging to analyze.

Based on the effective $f$--$f$ hopping matrix on a $Z$-bond derived via the Slater-Koster approach in Eq.~(\ref{eq:hopping_parameters_with_ligand_SOC}), we systematically categorize the nonzero indirect $f$--$p$--$f$ hopping channels by their Pauli matrix components, as illustrated in Fig.~\ref{fig:hopping_channels}. After analyzing the effect of the ligand-SOC on these hopping channels, we classify them into two types:

\textbf{Type A}: The intrinsic indirect hopping channels originating from the $90^\circ$ Sm--I--Sm bonding geometry, characterized by the $\sigma^0$ components. Comparative analysis of Eqs.~(\ref{eq:hopping_parameters_with_ligand_SOC}) and (\ref{eq:hopping_parameters_without_ligand_SOC}) demonstrates that the ligand-SOC modifies the energy levels of the ligand $5p$ spin-orbit-entangled states, introducing an additional scaling factor $\frac{\Delta}{3}(\frac{2}{\Delta-\lambda_p/2}+\frac{1}{\Delta+\lambda_p})$ to the effective $f$--$f$ indirect hoppings.

\textbf{Type B}: The ligand-SOC-activated hopping channels associated with the onsite hoppings between the spin-orbit-entangled states at the ligand sites, characterized by the $\sigma^x,\sigma^y,\sigma^z$ components. These can be further categorized into spin-flip (SF) channels ($\sigma^x,\sigma^y$) and spin-conserving (SC) channels ($\sigma^z$). Such channels are otherwise forbidden due to the simultaneous conservation of orbital angular momentum $l$ and spin angular momentum $s$ during indirect hoppings. However, the ligand-SOC couples $l$ and $s$ into the total angular momentum $j=l+s$, enabling these channels under $j$ conservation. Notably, these ligand-SOC-activated channels make significant contributions to the AFM Kitaev interactions, as shown in  Fig.~\ref{fig:reorder_contributions_J_K}. 

\begin{figure}[!htbp]
\includegraphics[width=0.48\textwidth]{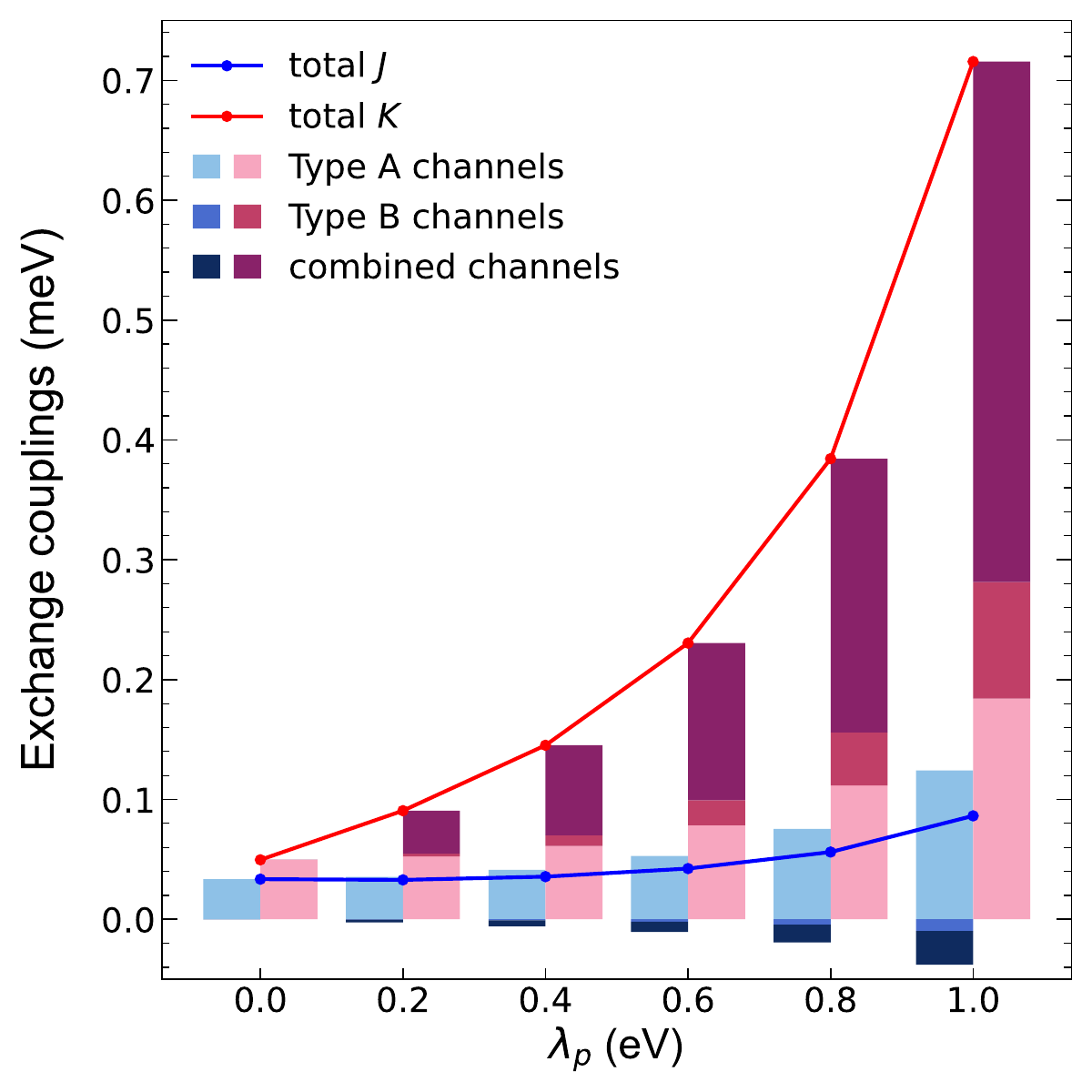}
\caption{
Contributions to the Heisenberg coupling $J$ and Kitaev coupling $K$ from different types of exchange channels as a function of the ligand-SOC strength $\lambda_p$, calculated using hopping parameters derived from the Slater-Koster approach. The Slater-Koster integrals are set to $t_{pf\sigma}=0.24\ \text{eV}$ and $t_{pf\pi}/t_{pf\sigma}=-0.5$, with the $4f$-$5p$ atomic energy difference $\Delta$ estimated as $0.8\ \text{eV}$. The onsite Coulomb energy $U$,  Hund’s coupling ratio $J_h/U$, atomic SOC strength of $\text{Sm}^{3+}$ $\lambda$ and crystal field parameter $B_4^0$ are set to be 5.53 eV, 0.15, 36.9 $\text{meV}$ and $0.048$ meV, respectively, consistent with earlier sections. The light blue (red), blue (red) and dark blue (red) bars represent the contributions to $J$ ($K$) from the individual Type A, Type B and combined Type A + Type B channels, respectively. The blue and red lines indicate the total $J$ and $K$ as functions of $\lambda_p$. 
}
\label{fig:SK_switch}
\end{figure}

We also compute the contributions of these two types of channels to the exchange interactions as the ligand-SOC strength increases. The results are shown in Fig.~\ref{fig:SK_switch}. As the ligand-SOC strength $\lambda_p$ increases, the contributions to the AFM Kitaev coupling $K$ from the individual Type A and Type B channels show a moderate increase. However, that contributions from the combined Type A $+$ Type B channels exhibit significant growth, eventually dominating the AFM Kitaev coupling. For the Heisenberg coupling $J$, both the Type B and combined Type A $+$ Type B channels yield weak FM contributions, which partially offset the AFM contributions from the individual Type A channels. As a result, with increasing $\lambda_p$, the Kitaev coupling $K$ ultimately becomes the dominant exchange term. 
These findings highlight the essential role of the ligand-SOC-activated Type B hopping channels---particularly their synergy with the intrinsic Type A channels---in driving the emergence of the dominant AFM Kitaev interactions in $\text{SmI}_3$.

\subsection{Spin excitation spectra}
\label{subsec:Spin excitation spectra}

\begin{figure}[!htbp]
\includegraphics[width=0.5\textwidth]{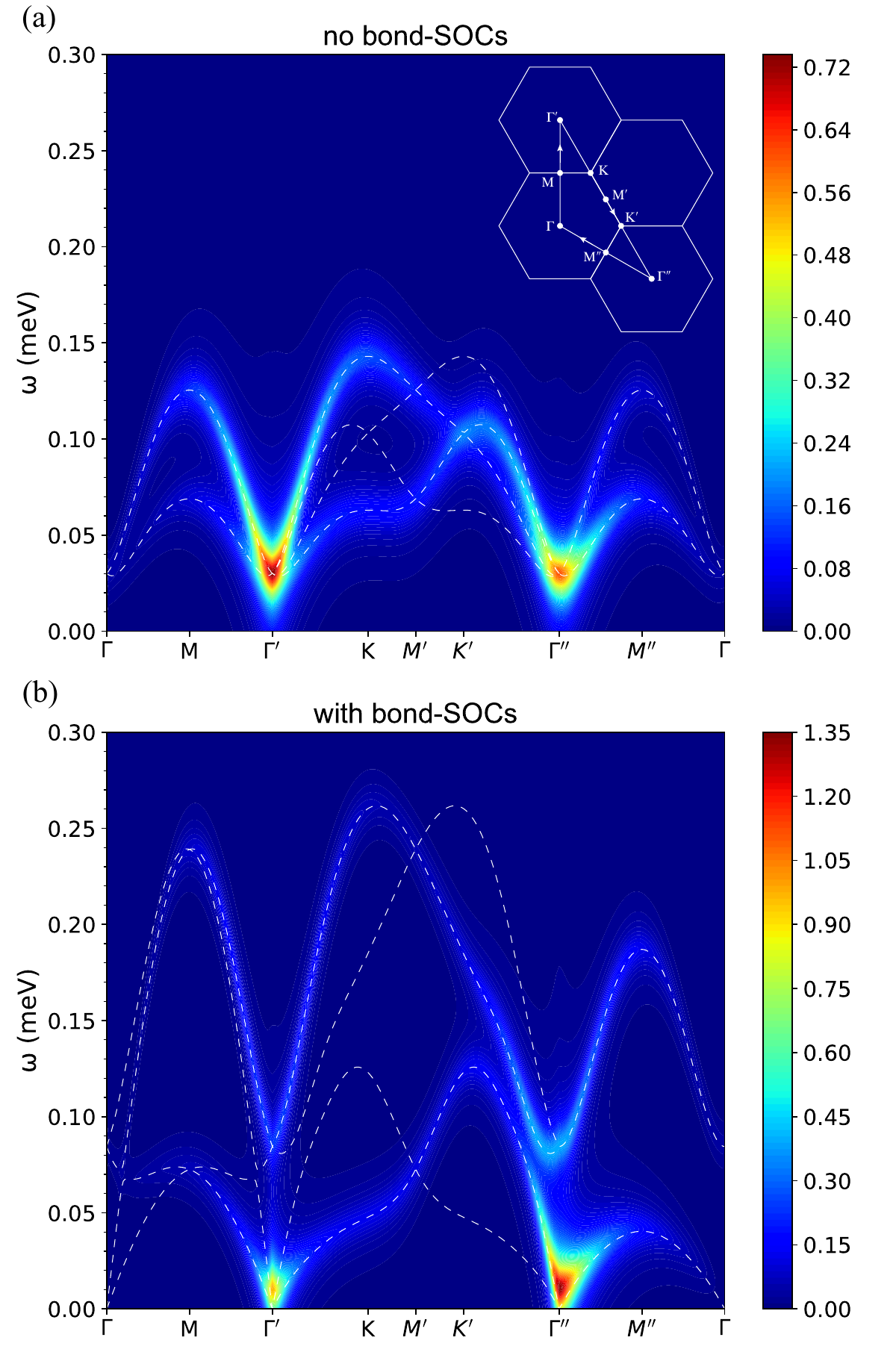}
\caption{
LSWT spectra of the model using the two sets of parameters from Table~\ref{tab:exchange parameters}.  
(a) Spectra for parameters without the bond-SOCs, based on an AFM order along the $[1,1,1]$-direction. 
(b) Spectra for parameters with the bond-SOCs, based on an AFM order along the $[1,1,-2]$-direction. The inset shows the high symmetry path in reciprocal space.}
\label{fig:spin_wave}
\end{figure}

We further proceed on the AFM orders and spin excitations of $\text{SmI}_3$ for both sets of exchange parameters from Table~\ref{tab:exchange parameters} to elucidate the effect of the bond-SOCs. Although both cases lie in the AFM region, the one with the bond-SOCs is much closer to the Kitaev-QSL phase. A 24-site exact diagonalization shows that the ground state AFM order without the bond-SOCs is oriented perpendicular to the plane along the $[1,1,1]$-direction, whereas the AFM order with the bond-SOCs lies within the plane. Given the $C_3$ rotational symmetry of the honeycomb lattice, this in-plane order breaks symmetry due to the enhanced fluctuations near the AFM Kitaev point induced by the ligand-SOC. Furthermore, the breaking of the $C_3$ symmetry is expected to affect the magnon dispersion.

To analyze the spin excitations, we employ linear spin-wave theory (LSWT) for both AFM orders \cite{GeneralizedSpinwave_Muniz_2014,SUSpinwave_Dong_2018}. As shown in Fig.~\ref{fig:spin_wave}~(a), since the $[1,1,1]$-direction AFM order preserves all model symmetries, a clear gap appears in the LSWT spectra without the bond-SOCs due to the anisotropy of the model. The high intensity located at $\Gamma'$ point indicates the stability of this AFM order. When the bond-SOCs are included, the AFM order flops into the plane; here, we choose the $[1,1,-2]$-direction order to break the $C_3$ rotational symmetry. The corresponding LSWT spectra, shown in Fig.~\ref{fig:spin_wave}~(b), reveals the emergence of gapless modes, similar to Goldstone modes. Since the $C_3$ symmetry is effectively restored to a continuous $U(1)$ symmetry within the LSWT approximation, the breaking of $C_3$ symmetry mimics the breaking of $U(1)$ symmetry, leading to these gapless modes. Therefore, such modes are expected to be unstable under high-order perturbations. It is noted that the in-plane order direction cannot be determined from exact diagonalization and remains classically degenerate within the LSWT approximation. Consequently, the spin-wave dispersion varies depending on the chosen order direction.

\begin{figure}[!htbp]
\includegraphics[width=0.4\textwidth]{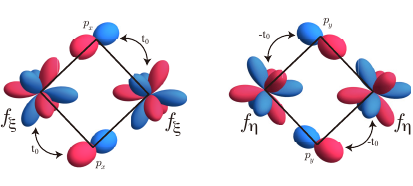}
\caption{
Schematic representation of two distortion-enhanced hopping channels $\xi\xi$ and $\eta\eta$. Under the assumption of an ideal $90^\circ$ bonding geometry, these indirect $f$--$p$--$f$ hopping parameters are exactly zero within the Slater-Koster framework. However, lattice distortions lift this restriction and induce finite amplitudes for these channels. The inclusion of the ligand-SOC can further enhance their strength by introducing an additional scaling factor.
}
\label{fig:extra_channels}
\end{figure}

\begin{figure*}[!htbp]
\includegraphics[width=\textwidth]{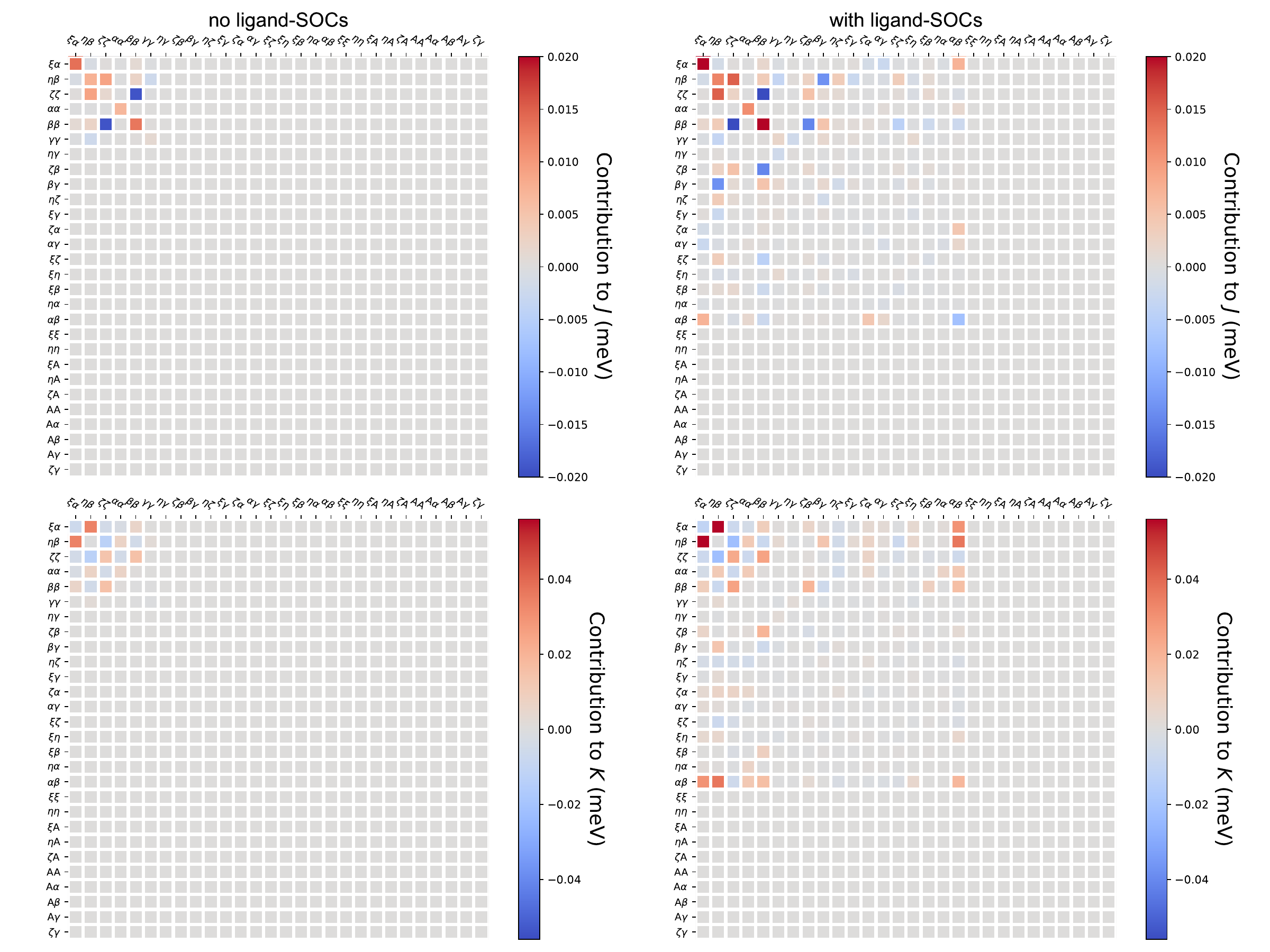}
\caption{ 
Heatmaps illustrating the contributions from various exchange channels to the Heisenberg coupling $J$ and the Kitaev coupling $K$ on a $Z$-bond, with and without ligand-SOCs, using the Slater-Koster approach. The label pair $(uv;u'v')$ indicates the exchange channel composed of hopping parameters $t_{iu\sigma,jv\sigma'}$ and $t_{iu'\sigma,jv'\sigma'}$ on a $Z$-bond, where $u,v,u',v'\in \{\xi,\eta,\zeta,A,\alpha,\beta,\gamma$\} and $\sigma,\sigma' \in \{ \uparrow,\downarrow \}$. Red units indicate AFM contributions, while blue units correspond to FM contributions. The onsite Coulomb energy  is set to $U=5.53$ eV, with a Hund’s coupling  ratio of $J_h/U=0.15$. The atomic SOC strength of $\text{Sm}^{3+}$ and the crystal field parameter are set to $\lambda=36.9\ \text{meV}$ and $B_4^0=4.8\times10^{-2}\ \text{meV}$, respectively. The Slater-Koster integrals are chosen as $t_{pf\sigma}=0.24\ \text{eV}$, $t_{pf\pi}/t_{pf\sigma}=-0.5$ with $\lambda_p=0.63\ \text{eV}$ and $\Delta=0.8\ \text{eV}$, to fit the DFT results.
}
\label{fig:reorder_SK_contributions_J_K}
\end{figure*}

\section{discussion and summary}
\label{sec:discussion and summary}

In Sec.~\ref{subsec:effect of ligand-SOC}, we investigate the role of the ligand-SOC in exchange processes based on the Slater-Koster analysis, derived through a perturbative procedure assuming an ideal $90^\circ$ bonding geometry and neglecting direct $f$--$f$ hoppings. Comparing the hopping parameters obtained from the Slater-Koster approach (Eq.~(\ref{eq:hopping_parameters_with_ligand_SOC})) with those extracted from our DFT calculations (Tables.~\ref{tab:hopping parameters} and~\ref{tab:relativistic hopping parameters}), we find good agreement in the leading terms, except for three channels: $\xi\xi$, $\eta\eta$, and $\gamma\gamma$. The DFT-calculated hopping parameters for these channels are significantly larger than those predicted by the Slater-Koster approach. In fact, under the ideal $90^\circ$ bonding geometry assumption, the ligand-mediated indirect hoppings for $\xi\xi$ and $\eta\eta$ are strictly forbidden within the Slater-Koster framework (see Fig.~\ref{fig:extra_channels}). Interestingly, all three of these enhanced hopping channels belong to the $\sigma^0$ type, indicating no spin or orbital transfer occurs at the ligand sites. We suggest that these increased hopping amplitudes arise from the lattice distortions, which deviate from the ideal $90^\circ$ bonding geometry and are captured in our DFT calculations. To assess their impact on the exchange interactions, we also compute the contributions of different exchange channels using the Slater-Koster hopping parameters for comparison. The results, shown in Fig.~\ref{fig:reorder_SK_contributions_J_K}, can be directly compared with those in Fig.~\ref{fig:reorder_contributions_J_K}, revealing that these distortion-enhanced channels make substantial contributions, particularly to the AFM Heisenberg coupling $J$. This observation suggests that fine-tuning lattice distortions could provide an effective pathway to modulate $J$ and stabilize the ideal Kitaev-QSL phase in $\text{SmI}_3$.

In summary, we have systematically investigated the effective exchange interactions and magnetic properties of the $4f^5$ honeycomb iodide $\text{SmI}_3$. Starting from the onsite multi-orbital Hamiltonian, we construct both the low- and high-energy Hilbert spaces and confirm that the ground states correspond to the $\Gamma_7$ Kramers doublet over a wide parameter range, consistent with experimental observations. To determine the band structure and hopping parameters, we perform \textit{ab initio} DFT calculations using experimentally measured structural data. 
By selectively enabling and disabling SOCs in our DFT calculations, we identify the bond-dependent SOC effects in the hopping processes—termed bond-SOCs.
Through the numerical strong coupling expansion, we derive the minimal nearest-neighbor pseudospin exchange model. Our results demonstrate that the bond-SOCs significantly enhance the AFM Kitaev interaction relative to other exchange terms, leading to a dominant AFM Kitaev coupling. A detailed analysis of the exchange channels reveals that nearly all additional channels activated by the bond-SOCs contribute to the AFM Kitaev interaction. To elucidate the microscopic origin of the bond-SOCs and the emergence of new hopping channels, we employ the Slater-Koster approach, incorporating the strong SOC of the heavy iodine ligands. The resulting effective $f$--$f$ hopping parameters closely resemble those from our DFT calculations, confirming that the bond-SOCs originate from the strong ligand-SOC. We further categorize the hopping
channels into two types: the intrinsic channels (Type A) and new ligand-SOC-activated channels (Type B). Notably, the combined exchange channels (Type A $+$ Type B) play a crucial role in driving the dominant AFM Kitaev interaction. 
For the AFM orders, a 24-site exact diagonalization reveals a spin-flop transition induced by the bond-SOCs, shifting the AFM order from the out-of-plane $[1, 1, 1]$-direction to an in-plane orientation, breaking $C_3$ rotational symmetry. The emergence of gapless modes in the spin excitation spectral suggests enhanced quantum fluctuations and instability near the AFM Kitaev point. 
In conclusion, our findings highlight the crucial role of ligand-SOC in stabilizing the dominant AFM Kitaev interactions in the $4f^5$ honeycomb iodide $\text{SmI}_3$,  offering valuable insights for the exploration of new $f$-electron Kitaev-QSL candidates. Further experimental studies, such as angle-dependent ferromagnetic resonance (FMR) and inelastic neutron scattering (INS), are essential to determine the strength of the exchange interactions and to identify the excitation signatures of Kitaev-QSL phases. 

We also take note of a recent work by Jang and Motome \cite{jang_exploring_2024}, in which they systematically examined all possible $4f$ electron conﬁgurations using the Slater-Koster approach and identified $4f^{3}$ and $4f^{11}$ as the most promising candidates for realizing the Kitaev interactions. Using the same parameters, we compute the $4f^5$ configuration while setting the ligand-SOC $\lambda_p$ to zero and find good agreement with their results. Upon introducing a finite ligand-SOC of $\lambda_p=0.63\ \text{eV}$, we observe a significant enhancement of the AFM Kitaev coupling $K$, with the ratio $|K/J|$ increasing from $2.5$ to $4.8$. This finding confirms the crucial role of the strong ligand-SOC in stabilizing dominant AFM Kitaev interactions and opens new possibilities for exploring similar effects in other $4f$ electron systems they proposed with heavy ligands, such as $4f^3$ $\text{NdI}_3$ and $4f^{11}$ $\text{ErI}_3$. Further theoretical investigations---particularly those based on \textit{ab initio} calculations--- alongside experimental verification, will be essential to assess their potential as Kitaev-QSL candidates.

\begin{acknowledgments}
The work was supported by National Key Projects for Research and Development of China (Grant No. 2021YFA1400400), the National Natural Science Foundation of China (Grant No. 92165205 and No. 12434005). We thank e-Science Center of Collaborative Innovation Center of Advanced Microstructures for providing computational resources.
\end{acknowledgments}

\appendix

\section{\textit{\bf {\it ab initio}} RESULTS}
\label{appendix:ab initio results}

\begin{figure*}[!htbp]
\includegraphics[width=0.8\textwidth]{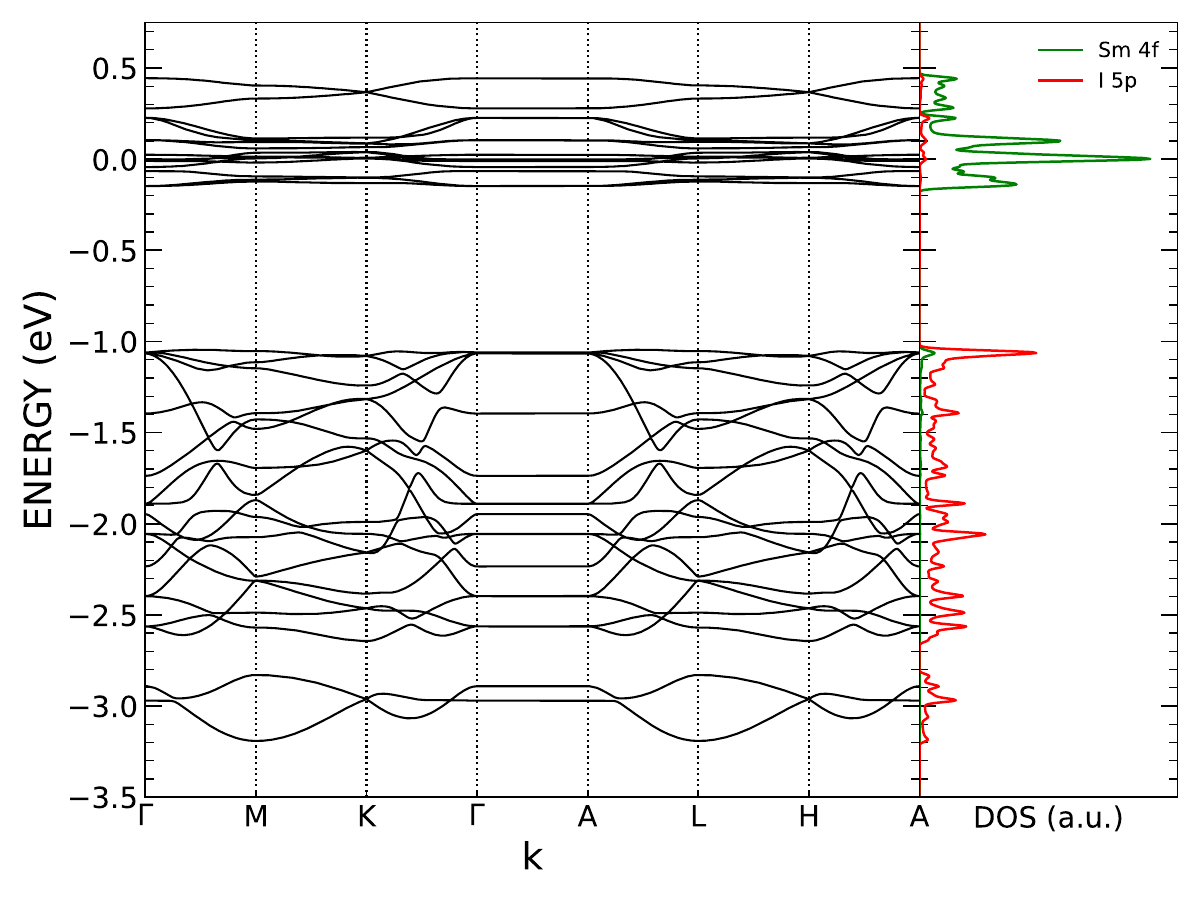}
\caption{ 
Electronic band structures and the projected density of states for $\text{Sm}$ $4f$ electrons and $\text{I}$ $5p$ electrons without SOCs. The Fermi level is set to zero.}
\label{fig:pdos}
\end{figure*}

\begin{figure*}[!htbp]
\includegraphics[width=\textwidth]{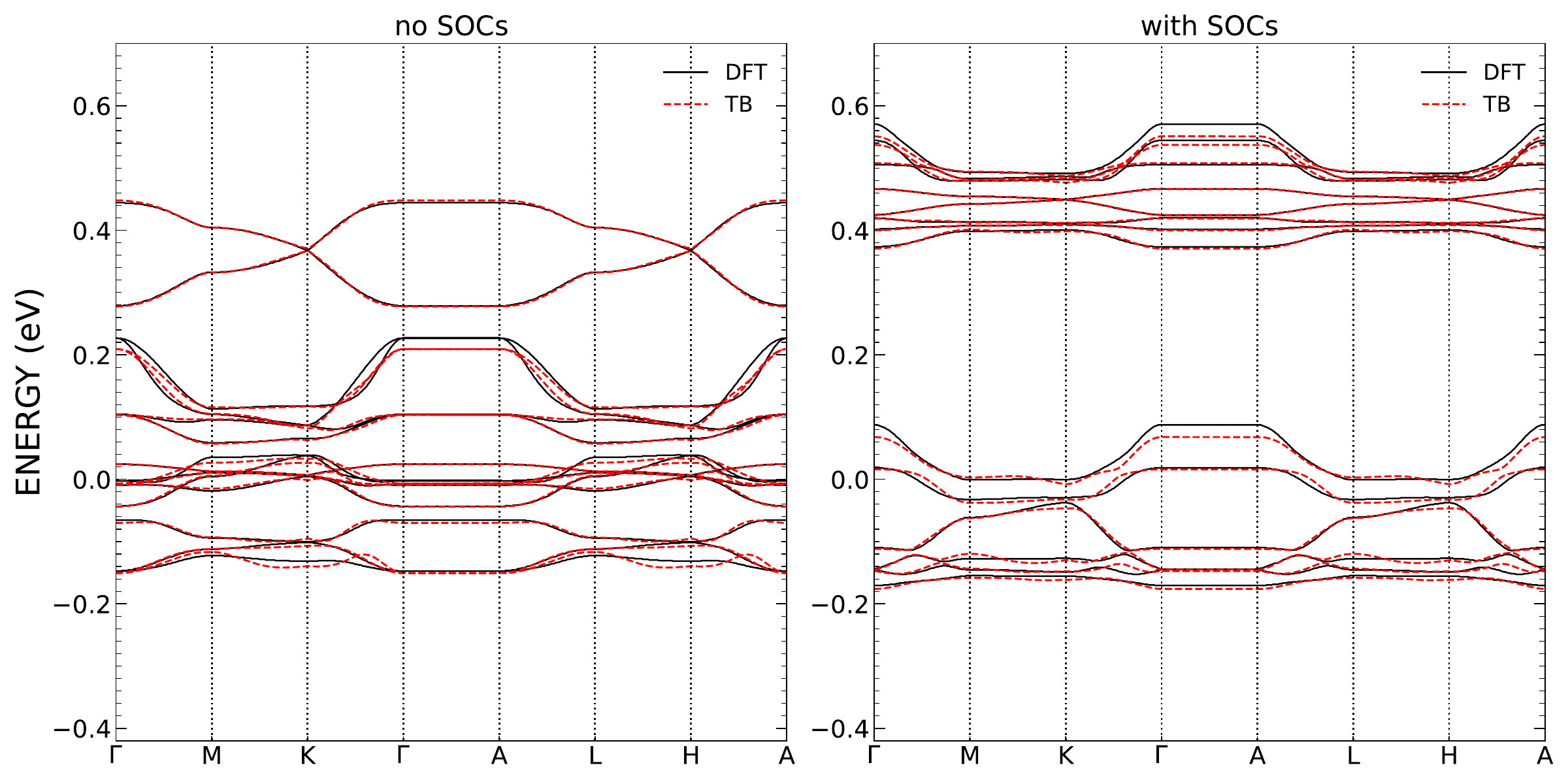}
\caption{ 
Electronic band structures focusing on $\text{Sm}$ $4f$ bands without and with SOCs. The black solid lines and red dashed lines show the band dispersions obtained by the DFT calculations and the tight-binding calculations with nearest-neighbor transfer integrals estimated by the MLWFs, respectively. The Fermi level is set to zero.}
\label{fig:DFT}
\end{figure*}

To determine the band structures and hopping parameters of $\text{SmI}_3$, we performed \textit{ab initio} calculations using the structural parameters obtained from experimental data \cite{ishikawa_SmI3_2022}.
Fig.~\ref{fig:pdos} prsents the band structures and projected density of states (PDOS) for $\text{Sm}$ $4f$ and $\text{I}$ $5p$ electrons, calculated via the density functional theory (DFT) without SOCs. The $\text{Sm}$ $4f$ bands are primarily located within the energy range of $-0.2$ eV to $0.5$ eV, exhibiting minimal hybridization with the $\text{I}$ $5p$ bands, which extend from $-3.2$ eV to $-1.0$ eV. This clear separation between the $4f$ and $5p$ bands ensures the validity of the maximally localized Wannier functions (MLWFs) method for extracting hopping parameters and constructing the effective exchange model. 
By selectively enabling
and disabling the SOCs in our DFT calculations, we obtained two sets of $\text{Sm}$ $4f$ bands, as shown in Fig.~\ref{fig:DFT} (black solid lines). The tight-binding (TB) band structures, derived using the nearest-neighbor transfer integrals estimated from MLWFs, are also plotted (red dashed lines), which closely match the DFT results.

\section{NEAREST-NEIGHBOR HOPPING PARAMETERS}
\label{appendix:hopping parameters}

\begin{table*}
\caption{
Nearest-neighbor hopping integrals on a $Z$-bond in the cubic basis, extracted from DFT calculations without SOCs. The unit is meV. Only the lower-left half of the table is shown, as the matrix is Hermitian.}
\label{tab:hopping parameters}
\begin{ruledtabular}
\begin{tabular}{cccccccccc}
  &$Z$-bond &$\xi$ &$\eta$ &$\zeta$ &$A$ &$\alpha$ &$\beta$ &$\gamma$\\ \hline
  &$\xi$ &16.4 & & & & & & \\
  &$\eta$ &11.7 &16.4 & & & & & \\
  &$\zeta$ &2.0 &2.0 &-22.5 & & & & \\
  &$A$ &-0.2 &0.2 &0.0 &-5.0 & & & \\
  &$\alpha$ &-26.9 &-7.9 &-2.5 &-3.4 &36.8 & & \\
  &$\beta$ &7.9 &26.9 &2.5 &-3.4 &-8.6 &36.8 & \\
  &$\gamma$ &7.8 &-7.8 &0.0 &-5.8 &-1.3 &-1.3 &22.6 \\
 \end{tabular}
\end{ruledtabular}
\end{table*}

\begin{table*}
\caption{
Bond-SOC vector matrix $\vec{\lambda}$ (as defined in Eq.~(\ref{eq:bond-SOCs})) on a $Z$-bond in the cubic basis. The unit is meV. Only the lower-left half of the table is shown, as the matrix is Hermitian. The dominant terms are highlighted in bold.}
\label{tab:relativistic hopping parameters}
\begin{ruledtabular}
\begin{tabular}{cccccccccc}
  &$\lambda^0$ &$\xi$ &$\eta$ &$\zeta$ &$A$ &$\alpha$ &$\beta$ &$\gamma$\\ \hline
  &$\xi$ &\textbf{10.6} & & & & & & \\
  &$\eta$ &-2.4 &\textbf{10.6} & & & & & \\
  &$\zeta$ &-0.6 &-0.6 &-3.8 & & & & \\
  &$A$ &0.6 &-0.6 &0.0 &-1.0 & & & \\
  &$\alpha$ &-4.6 &5.0 &3.0 &1.8 &-5.2 & & \\
  &$\beta$ &-5.0 &4.6 &-3.0 &1.8 &6.2 &-5.2 & \\
  &$\gamma$ &-7.0 &7.0 &0.0 &-2.2 &4.0 &4.0 &\textbf{18.2} \\
  \hline\hline
  &$\lambda^x$ &$\xi$ &$\eta$ &$\zeta$ &$A$ &$\alpha$ &$\beta$ &$\gamma$\\ \hline
  &$\xi$ &0.0i & & & & & & \\
  &$\eta$ &2.0i &0.0i & & & & & \\
  &$\zeta$ &-1.8i &-2.8i &0.0i & & & & \\
  &$A$ &2.8i &-0.4i &0.2i &0.0i & & & \\
  &$\alpha$ &-0.6i &-0.2i &-1.2i &3.4i &0.0i & & \\
  &$\beta$ &4.4i &2.2i &\textbf{10.8i} &-3.4i &-0.6i &0.0i & \\
  &$\gamma$ &0.8i &\textbf{22.6i} &-2.4i &-2.4i &-5.6i &\textbf{10.0i} &0.0i \\
  \hline\hline
  &$\lambda^y$ &$\xi$ &$\eta$ &$\zeta$ &$A$ &$\alpha$ &$\beta$ &$\gamma$\\ \hline
  &$\xi$ &0.0i & & & & & & \\
  &$\eta$ &2.0i &0.0i & & & & & \\
  &$\zeta$ &2.8i &1.8i &0.0i & & & & \\
  &$A$ &-0.4i &2.8i &0.2i &0.0i & & & \\
  &$\alpha$ &2.2i &4.4i &\textbf{10.8i} &3.4i &0.0i & & \\
  &$\beta$ &-0.2i &-0.6i &-1.2i &-3.4i &-0.6i &0.0i & \\
  &$\gamma$ &\textbf{22.6i} &0.8i &-2.4i &2.4i &\textbf{-10.0i} &5.6i &0.0i \\
  \hline\hline
  &$\lambda^z$ &$\xi$ &$\eta$ &$\zeta$ &$A$ &$\alpha$ &$\beta$ &$\gamma$\\ \hline
  &$\xi$ &0.0i & & & & & & \\
  &$\eta$ &\textbf{-20.0i} &0.0i & & & & & \\
  &$\zeta$ &-0.6i &0.6i &0.0i & & & & \\
  &$A$ &0.2i &0.2i &-1.2i &0.0i & & & \\
  &$\alpha$ &3.6i &\textbf{-19.6i} &-0.2i &-0.6i &0.0i & & \\
  &$\beta$ &\textbf{-19.6i} &3.6i &-0.2i &0.6i &\textbf{16.2i} &0.0i & \\
  &$\gamma$ &2.4i &2.4i &-1.2i &0.0i &-3.8i &3.8i &0.0i \\
 \end{tabular}
\end{ruledtabular}
\end{table*}

\bibliography{main.bib}

\begin{thebibliography}{75}%
\makeatletter
\providecommand \@ifxundefined [1]{%
 \@ifx{#1\undefined}
}%
\providecommand \@ifnum [1]{%
 \ifnum #1\expandafter \@firstoftwo
 \else \expandafter \@secondoftwo
 \fi
}%
\providecommand \@ifx [1]{%
 \ifx #1\expandafter \@firstoftwo
 \else \expandafter \@secondoftwo
 \fi
}%
\providecommand \natexlab [1]{#1}%
\providecommand \enquote  [1]{``#1''}%
\providecommand \bibnamefont  [1]{#1}%
\providecommand \bibfnamefont [1]{#1}%
\providecommand \citenamefont [1]{#1}%
\providecommand \href@noop [0]{\@secondoftwo}%
\providecommand \href [0]{\begingroup \@sanitize@url \@href}%
\providecommand \@href[1]{\@@startlink{#1}\@@href}%
\providecommand \@@href[1]{\endgroup#1\@@endlink}%
\providecommand \@sanitize@url [0]{\catcode `\\12\catcode `\$12\catcode
  `\&12\catcode `\#12\catcode `\^12\catcode `\_12\catcode `\%12\relax}%
\providecommand \@@startlink[1]{}%
\providecommand \@@endlink[0]{}%
\providecommand \url  [0]{\begingroup\@sanitize@url \@url }%
\providecommand \@url [1]{\endgroup\@href {#1}{\urlprefix }}%
\providecommand \urlprefix  [0]{URL }%
\providecommand \Eprint [0]{\href }%
\providecommand \doibase [0]{https://doi.org/}%
\providecommand \selectlanguage [0]{\@gobble}%
\providecommand \bibinfo  [0]{\@secondoftwo}%
\providecommand \bibfield  [0]{\@secondoftwo}%
\providecommand \translation [1]{[#1]}%
\providecommand \BibitemOpen [0]{}%
\providecommand \bibitemStop [0]{}%
\providecommand \bibitemNoStop [0]{.\EOS\space}%
\providecommand \EOS [0]{\spacefactor3000\relax}%
\providecommand \BibitemShut  [1]{\csname bibitem#1\endcsname}%
\let\auto@bib@innerbib\@empty
%</preamble>
\bibitem [{\citenamefont {Anderson}(1973)}]{anderson_resonating_1973}%
  \BibitemOpen
  \bibfield  {author} {\bibinfo {author} {\bibfnamefont {P.}~\bibnamefont
  {Anderson}},\ }\bibfield  {title} {\bibinfo {title} {Resonating valence
  bonds: {A} new kind of insulator?},\ }\href
  {https://doi.org/10.1016/0025-5408(73)90167-0} {\bibfield  {journal}
  {\bibinfo  {journal} {Materials Research Bulletin}\ }\textbf {\bibinfo
  {volume} {8}},\ \bibinfo {pages} {153} (\bibinfo {year} {1973})}\BibitemShut
  {NoStop}%
\bibitem [{\citenamefont {Balents}(2010)}]{balents_spin_2010}%
  \BibitemOpen
  \bibfield  {author} {\bibinfo {author} {\bibfnamefont {L.}~\bibnamefont
  {Balents}},\ }\bibfield  {title} {\bibinfo {title} {Spin liquids in
  frustrated magnets},\ }\href {https://doi.org/10.1038/nature08917} {\bibfield
   {journal} {\bibinfo  {journal} {Nature}\ }\textbf {\bibinfo {volume}
  {464}},\ \bibinfo {pages} {199} (\bibinfo {year} {2010})}\BibitemShut
  {NoStop}%
\bibitem [{\citenamefont {Savary}\ and\ \citenamefont
  {Balents}(2016)}]{savary_quantum_2016}%
  \BibitemOpen
  \bibfield  {author} {\bibinfo {author} {\bibfnamefont {L.}~\bibnamefont
  {Savary}}\ and\ \bibinfo {author} {\bibfnamefont {L.}~\bibnamefont
  {Balents}},\ }\bibfield  {title} {\bibinfo {title} {Quantum spin liquids: a
  review},\ }\href {https://doi.org/10.1088/0034-4885/80/1/016502} {\bibfield
  {journal} {\bibinfo  {journal} {Reports on Progress in Physics}\ }\textbf
  {\bibinfo {volume} {80}},\ \bibinfo {pages} {016502} (\bibinfo {year}
  {2016})}\BibitemShut {NoStop}%
\bibitem [{\citenamefont {Zhou}\ \emph {et~al.}(2017)\citenamefont {Zhou},
  \citenamefont {Kanoda},\ and\ \citenamefont {Ng}}]{zhou_quantum_2017}%
  \BibitemOpen
  \bibfield  {author} {\bibinfo {author} {\bibfnamefont {Y.}~\bibnamefont
  {Zhou}}, \bibinfo {author} {\bibfnamefont {K.}~\bibnamefont {Kanoda}},\ and\
  \bibinfo {author} {\bibfnamefont {T.-K.}\ \bibnamefont {Ng}},\ }\bibfield
  {title} {\bibinfo {title} {Quantum spin liquid states},\ }\href
  {https://doi.org/10.1103/RevModPhys.89.025003} {\bibfield  {journal}
  {\bibinfo  {journal} {Rev. Mod. Phys.}\ }\textbf {\bibinfo {volume} {89}},\
  \bibinfo {pages} {025003} (\bibinfo {year} {2017})}\BibitemShut {NoStop}%
\bibitem [{\citenamefont {Wen}\ \emph {et~al.}(2019)\citenamefont {Wen},
  \citenamefont {Yu}, \citenamefont {Li}, \citenamefont {Yu},\ and\
  \citenamefont {Li}}]{ExperimentalIdentification_Wen_2019}%
  \BibitemOpen
  \bibfield  {author} {\bibinfo {author} {\bibfnamefont {J.}~\bibnamefont
  {Wen}}, \bibinfo {author} {\bibfnamefont {S.-L.}\ \bibnamefont {Yu}},
  \bibinfo {author} {\bibfnamefont {S.}~\bibnamefont {Li}}, \bibinfo {author}
  {\bibfnamefont {W.}~\bibnamefont {Yu}},\ and\ \bibinfo {author}
  {\bibfnamefont {J.-X.}\ \bibnamefont {Li}},\ }\bibfield  {title} {\bibinfo
  {title} {Experimental identification of quantum spin liquids},\ }\href
  {https://doi.org/10.1038/s41535-019-0151-6} {\bibfield  {journal} {\bibinfo
  {journal} {npj Quantum Materials}\ }\textbf {\bibinfo {volume} {4}},\
  \bibinfo {pages} {12} (\bibinfo {year} {2019})}\BibitemShut {NoStop}%
\bibitem [{\citenamefont {Broholm}\ \emph {et~al.}(2020)\citenamefont
  {Broholm}, \citenamefont {Cava}, \citenamefont {Kivelson}, \citenamefont
  {Nocera}, \citenamefont {Norman},\ and\ \citenamefont
  {Senthil}}]{QuantumSpin_Broholm_2020}%
  \BibitemOpen
  \bibfield  {author} {\bibinfo {author} {\bibfnamefont {C.}~\bibnamefont
  {Broholm}}, \bibinfo {author} {\bibfnamefont {R.~J.}\ \bibnamefont {Cava}},
  \bibinfo {author} {\bibfnamefont {S.~A.}\ \bibnamefont {Kivelson}}, \bibinfo
  {author} {\bibfnamefont {D.~G.}\ \bibnamefont {Nocera}}, \bibinfo {author}
  {\bibfnamefont {M.~R.}\ \bibnamefont {Norman}},\ and\ \bibinfo {author}
  {\bibfnamefont {T.}~\bibnamefont {Senthil}},\ }\bibfield  {title} {\bibinfo
  {title} {Quantum spin liquids},\ }\href
  {https://doi.org/10.1126/science.aay0668} {\bibfield  {journal} {\bibinfo
  {journal} {Science}\ }\textbf {\bibinfo {volume} {367}},\ \bibinfo {pages}
  {eaay0668} (\bibinfo {year} {2020})}\BibitemShut {NoStop}%
\bibitem [{\citenamefont {Matsuda}\ \emph {et~al.}(2025)\citenamefont
  {Matsuda}, \citenamefont {Shibauchi},\ and\ \citenamefont
  {Kee}}]{matsuda_KitaevQuantumSpin_2025}%
  \BibitemOpen
  \bibfield  {author} {\bibinfo {author} {\bibfnamefont {Y.}~\bibnamefont
  {Matsuda}}, \bibinfo {author} {\bibfnamefont {T.}~\bibnamefont {Shibauchi}},\
  and\ \bibinfo {author} {\bibfnamefont {H.-Y.}\ \bibnamefont {Kee}},\ }\href
  {https://doi.org/10.48550/arXiv.2501.05608} {\bibinfo {title} {Kitaev
  {{Quantum Spin Liquids}}}} (\bibinfo {year} {2025}),\ \Eprint
  {https://arxiv.org/abs/2501.05608} {arXiv:2501.05608 [cond-mat]} \BibitemShut
  {NoStop}%
\bibitem [{\citenamefont {WEN}(1991)}]{wen_topological_1991}%
  \BibitemOpen
  \bibfield  {author} {\bibinfo {author} {\bibfnamefont {X.-G.}\ \bibnamefont
  {WEN}},\ }\bibfield  {title} {\bibinfo {title} {{TOPOLOGICAL} {ORDERS} {AND}
  {CHERN}-{SIMONS} {THEORY} {IN} {STRONGLY} {CORRELATED} {QUANTUM} {LIQUID}},\
  }\href {https://doi.org/10.1142/S0217979291001541} {\bibfield  {journal}
  {\bibinfo  {journal} {International Journal of Modern Physics B}\ }\textbf
  {\bibinfo {volume} {05}},\ \bibinfo {pages} {1641} (\bibinfo {year}
  {1991})}\BibitemShut {NoStop}%
\bibitem [{\citenamefont {Wen}(2002)}]{QuantumOrders_Wen_2002a}%
  \BibitemOpen
  \bibfield  {author} {\bibinfo {author} {\bibfnamefont {X.-G.}\ \bibnamefont
  {Wen}},\ }\bibfield  {title} {\bibinfo {title} {Quantum orders and symmetric
  spin liquids},\ }\href {https://doi.org/10.1103/PhysRevB.65.165113}
  {\bibfield  {journal} {\bibinfo  {journal} {Physical Review B}\ }\textbf
  {\bibinfo {volume} {65}},\ \bibinfo {pages} {165113} (\bibinfo {year}
  {2002})}\BibitemShut {NoStop}%
\bibitem [{\citenamefont {Sachdev}(1992)}]{sachdev_kagome-_1992}%
  \BibitemOpen
  \bibfield  {author} {\bibinfo {author} {\bibfnamefont {S.}~\bibnamefont
  {Sachdev}},\ }\bibfield  {title} {\bibinfo {title} {Kagome´- and
  triangular-lattice {Heisenberg} antiferromagnets: {Ordering} from quantum
  fluctuations and quantum-disordered ground states with unconfined bosonic
  spinons},\ }\href {https://doi.org/10.1103/PhysRevB.45.12377} {\bibfield
  {journal} {\bibinfo  {journal} {Physical Review B}\ }\textbf {\bibinfo
  {volume} {45}},\ \bibinfo {pages} {12377} (\bibinfo {year}
  {1992})}\BibitemShut {NoStop}%
\bibitem [{\citenamefont {Kitaev}(2003)}]{kitaev_fault-tolerant_2003}%
  \BibitemOpen
  \bibfield  {author} {\bibinfo {author} {\bibfnamefont {A.}~\bibnamefont
  {Kitaev}},\ }\bibfield  {title} {\bibinfo {title} {Fault-tolerant quantum
  computation by anyons},\ }\href
  {https://doi.org/10.1016/S0003-4916(02)00018-0} {\bibfield  {journal}
  {\bibinfo  {journal} {Annals of Physics}\ }\textbf {\bibinfo {volume}
  {303}},\ \bibinfo {pages} {2} (\bibinfo {year} {2003})}\BibitemShut {NoStop}%
\bibitem [{\citenamefont {Kitaev}(2006)}]{kitaev_anyons_2006}%
  \BibitemOpen
  \bibfield  {author} {\bibinfo {author} {\bibfnamefont {A.}~\bibnamefont
  {Kitaev}},\ }\bibfield  {title} {\bibinfo {title} {Anyons in an exactly
  solved model and beyond},\ }\href {https://doi.org/10.1016/j.aop.2005.10.005}
  {\bibfield  {journal} {\bibinfo  {journal} {Annals of Physics}\ }\textbf
  {\bibinfo {volume} {321}},\ \bibinfo {pages} {2} (\bibinfo {year}
  {2006})}\BibitemShut {NoStop}%
\bibitem [{\citenamefont {Jackeli}\ and\ \citenamefont
  {Khaliullin}(2009)}]{jackeli_mott_2009}%
  \BibitemOpen
  \bibfield  {author} {\bibinfo {author} {\bibfnamefont {G.}~\bibnamefont
  {Jackeli}}\ and\ \bibinfo {author} {\bibfnamefont {G.}~\bibnamefont
  {Khaliullin}},\ }\bibfield  {title} {\bibinfo {title} {Mott {Insulators} in
  the {Strong} {Spin}-{Orbit} {Coupling} {Limit}: {From} {Heisenberg} to a
  {Quantum} {Compass} and {Kitaev} {Models}},\ }\href
  {https://doi.org/10.1103/PhysRevLett.102.017205} {\bibfield  {journal}
  {\bibinfo  {journal} {Physical Review Letters}\ }\textbf {\bibinfo {volume}
  {102}},\ \bibinfo {pages} {017205} (\bibinfo {year} {2009})}\BibitemShut
  {NoStop}%
\bibitem [{\citenamefont {Chaloupka}\ \emph {et~al.}(2010)\citenamefont
  {Chaloupka}, \citenamefont {Jackeli},\ and\ \citenamefont
  {Khaliullin}}]{KitaevHeisenbergModel_Chaloupka_2010}%
  \BibitemOpen
  \bibfield  {author} {\bibinfo {author} {\bibfnamefont {J.}~\bibnamefont
  {Chaloupka}}, \bibinfo {author} {\bibfnamefont {G.}~\bibnamefont {Jackeli}},\
  and\ \bibinfo {author} {\bibfnamefont {G.}~\bibnamefont {Khaliullin}},\
  }\bibfield  {title} {\bibinfo {title} {Kitaev-{{Heisenberg Model}} on a
  {{Honeycomb Lattice}}: {{Possible Exotic Phases}} in iridium oxides
  $\text{A}_2\text{IrO}_3$},\ }\href
  {https://doi.org/10.1103/PhysRevLett.105.027204} {\bibfield  {journal}
  {\bibinfo  {journal} {Physical Review Letters}\ }\textbf {\bibinfo {volume}
  {105}},\ \bibinfo {pages} {027204} (\bibinfo {year} {2010})}\BibitemShut
  {NoStop}%
\bibitem [{\citenamefont {Singh}\ and\ \citenamefont
  {Gegenwart}(2010)}]{singh_antiferromagnetic_2010}%
  \BibitemOpen
  \bibfield  {author} {\bibinfo {author} {\bibfnamefont {Y.}~\bibnamefont
  {Singh}}\ and\ \bibinfo {author} {\bibfnamefont {P.}~\bibnamefont
  {Gegenwart}},\ }\bibfield  {title} {\bibinfo {title} {Antiferromagnetic
  {Mott} insulating state in single crystals of the honeycomb lattice material
  {$\text{Na}_2\text{IrO}_3$}},\ }\href
  {https://doi.org/10.1103/PhysRevB.82.064412} {\bibfield  {journal} {\bibinfo
  {journal} {Physical Review B}\ }\textbf {\bibinfo {volume} {82}},\ \bibinfo
  {pages} {064412} (\bibinfo {year} {2010})}\BibitemShut {NoStop}%
\bibitem [{\citenamefont {Singh}\ \emph {et~al.}(2012)\citenamefont {Singh},
  \citenamefont {Manni}, \citenamefont {Reuther}, \citenamefont {Berlijn},
  \citenamefont {Thomale}, \citenamefont {Ku}, \citenamefont {Trebst},\ and\
  \citenamefont {Gegenwart}}]{singh_relevance_2012}%
  \BibitemOpen
  \bibfield  {author} {\bibinfo {author} {\bibfnamefont {Y.}~\bibnamefont
  {Singh}}, \bibinfo {author} {\bibfnamefont {S.}~\bibnamefont {Manni}},
  \bibinfo {author} {\bibfnamefont {J.}~\bibnamefont {Reuther}}, \bibinfo
  {author} {\bibfnamefont {T.}~\bibnamefont {Berlijn}}, \bibinfo {author}
  {\bibfnamefont {R.}~\bibnamefont {Thomale}}, \bibinfo {author} {\bibfnamefont
  {W.}~\bibnamefont {Ku}}, \bibinfo {author} {\bibfnamefont {S.}~\bibnamefont
  {Trebst}},\ and\ \bibinfo {author} {\bibfnamefont {P.}~\bibnamefont
  {Gegenwart}},\ }\bibfield  {title} {\bibinfo {title} {Relevance of the
  {Heisenberg}-{Kitaev} {Model} for the {Honeycomb} {Lattice} {Iridates}
  {$\text{A}_2\text{IrO}_3$}},\ }\href
  {https://doi.org/10.1103/PhysRevLett.108.127203} {\bibfield  {journal}
  {\bibinfo  {journal} {Physical Review Letters}\ }\textbf {\bibinfo {volume}
  {108}},\ \bibinfo {pages} {127203} (\bibinfo {year} {2012})}\BibitemShut
  {NoStop}%
\bibitem [{\citenamefont {Chaloupka}\ \emph {et~al.}(2013)\citenamefont
  {Chaloupka}, \citenamefont {Jackeli},\ and\ \citenamefont
  {Khaliullin}}]{ZigzagMagnetic_Chaloupka_2013}%
  \BibitemOpen
  \bibfield  {author} {\bibinfo {author} {\bibfnamefont {J.}~\bibnamefont
  {Chaloupka}}, \bibinfo {author} {\bibfnamefont {G.}~\bibnamefont {Jackeli}},\
  and\ \bibinfo {author} {\bibfnamefont {G.}~\bibnamefont {Khaliullin}},\
  }\bibfield  {title} {\bibinfo {title} {Zigzag {{Magnetic Order}} in the
  iridium oxide $\text{Na}_2\text{IrO}_3$},\ }\href
  {https://doi.org/10.1103/PhysRevLett.110.097204} {\bibfield  {journal}
  {\bibinfo  {journal} {Physical Review Letters}\ }\textbf {\bibinfo {volume}
  {110}},\ \bibinfo {pages} {097204} (\bibinfo {year} {2013})}\BibitemShut
  {NoStop}%
\bibitem [{\citenamefont {Modic}\ \emph {et~al.}(2014)\citenamefont {Modic},
  \citenamefont {Smidt}, \citenamefont {Kimchi}, \citenamefont {Breznay},
  \citenamefont {Biffin}, \citenamefont {Choi}, \citenamefont {Johnson},
  \citenamefont {Coldea}, \citenamefont {Watkins-Curry}, \citenamefont
  {McCandless}, \citenamefont {Chan}, \citenamefont {Gandara}, \citenamefont
  {Islam}, \citenamefont {Vishwanath}, \citenamefont {Shekhter}, \citenamefont
  {McDonald},\ and\ \citenamefont {Analytis}}]{modic_realization_2014}%
  \BibitemOpen
  \bibfield  {author} {\bibinfo {author} {\bibfnamefont {K.~A.}\ \bibnamefont
  {Modic}}, \bibinfo {author} {\bibfnamefont {T.~E.}\ \bibnamefont {Smidt}},
  \bibinfo {author} {\bibfnamefont {I.}~\bibnamefont {Kimchi}}, \bibinfo
  {author} {\bibfnamefont {N.~P.}\ \bibnamefont {Breznay}}, \bibinfo {author}
  {\bibfnamefont {A.}~\bibnamefont {Biffin}}, \bibinfo {author} {\bibfnamefont
  {S.}~\bibnamefont {Choi}}, \bibinfo {author} {\bibfnamefont {R.~D.}\
  \bibnamefont {Johnson}}, \bibinfo {author} {\bibfnamefont {R.}~\bibnamefont
  {Coldea}}, \bibinfo {author} {\bibfnamefont {P.}~\bibnamefont
  {Watkins-Curry}}, \bibinfo {author} {\bibfnamefont {G.~T.}\ \bibnamefont
  {McCandless}}, \bibinfo {author} {\bibfnamefont {J.~Y.}\ \bibnamefont
  {Chan}}, \bibinfo {author} {\bibfnamefont {F.}~\bibnamefont {Gandara}},
  \bibinfo {author} {\bibfnamefont {Z.}~\bibnamefont {Islam}}, \bibinfo
  {author} {\bibfnamefont {A.}~\bibnamefont {Vishwanath}}, \bibinfo {author}
  {\bibfnamefont {A.}~\bibnamefont {Shekhter}}, \bibinfo {author}
  {\bibfnamefont {R.~D.}\ \bibnamefont {McDonald}},\ and\ \bibinfo {author}
  {\bibfnamefont {J.~G.}\ \bibnamefont {Analytis}},\ }\bibfield  {title}
  {\bibinfo {title} {Realization of a three-dimensional spin–anisotropic
  harmonic honeycomb iridate},\ }\href {https://doi.org/10.1038/ncomms5203}
  {\bibfield  {journal} {\bibinfo  {journal} {Nature Communications}\ }\textbf
  {\bibinfo {volume} {5}},\ \bibinfo {pages} {4203} (\bibinfo {year}
  {2014})}\BibitemShut {NoStop}%
\bibitem [{\citenamefont {Yamaji}\ \emph {et~al.}(2014)\citenamefont {Yamaji},
  \citenamefont {Nomura}, \citenamefont {Kurita}, \citenamefont {Arita},\ and\
  \citenamefont {Imada}}]{FirstPrinciplesStudy_Yamaji_2014}%
  \BibitemOpen
  \bibfield  {author} {\bibinfo {author} {\bibfnamefont {Y.}~\bibnamefont
  {Yamaji}}, \bibinfo {author} {\bibfnamefont {Y.}~\bibnamefont {Nomura}},
  \bibinfo {author} {\bibfnamefont {M.}~\bibnamefont {Kurita}}, \bibinfo
  {author} {\bibfnamefont {R.}~\bibnamefont {Arita}},\ and\ \bibinfo {author}
  {\bibfnamefont {M.}~\bibnamefont {Imada}},\ }\bibfield  {title} {\bibinfo
  {title} {First-principles study of the honeycomb-lattice iridates
  {$\text{Na}_2$}{$\text{IrO}_3$} in the presence of strong spin-orbit
  interaction and electron correlations},\ }\href
  {https://doi.org/10.1103/PhysRevLett.113.107201} {\bibfield  {journal}
  {\bibinfo  {journal} {Physical Review Letters}\ }\textbf {\bibinfo {volume}
  {113}},\ \bibinfo {pages} {107201} (\bibinfo {year} {2014})}\BibitemShut
  {NoStop}%
\bibitem [{\citenamefont {Rau}\ \emph {et~al.}(2014)\citenamefont {Rau},
  \citenamefont {Lee},\ and\ \citenamefont {Kee}}]{rau_generic_2014}%
  \BibitemOpen
  \bibfield  {author} {\bibinfo {author} {\bibfnamefont {J.~G.}\ \bibnamefont
  {Rau}}, \bibinfo {author} {\bibfnamefont {E.~K.-H.}\ \bibnamefont {Lee}},\
  and\ \bibinfo {author} {\bibfnamefont {H.-Y.}\ \bibnamefont {Kee}},\
  }\bibfield  {title} {\bibinfo {title} {Generic {Spin} {Model} for the
  {Honeycomb} {Iridates} beyond the {Kitaev} {Limit}},\ }\href
  {https://doi.org/10.1103/PhysRevLett.112.077204} {\bibfield  {journal}
  {\bibinfo  {journal} {Physical Review Letters}\ }\textbf {\bibinfo {volume}
  {112}},\ \bibinfo {pages} {077204} (\bibinfo {year} {2014})}\BibitemShut
  {NoStop}%
\bibitem [{\citenamefont {Hwan~Chun}\ \emph {et~al.}(2015)\citenamefont
  {Hwan~Chun}, \citenamefont {Kim}, \citenamefont {Kim}, \citenamefont {Zheng},
  \citenamefont {Stoumpos}, \citenamefont {Malliakas}, \citenamefont
  {Mitchell}, \citenamefont {Mehlawat}, \citenamefont {Singh}, \citenamefont
  {Choi}, \citenamefont {Gog}, \citenamefont {Al-Zein}, \citenamefont {Sala},
  \citenamefont {Krisch}, \citenamefont {Chaloupka}, \citenamefont {Jackeli},
  \citenamefont {Khaliullin},\ and\ \citenamefont
  {Kim}}]{hwan_chun_direct_2015}%
  \BibitemOpen
  \bibfield  {author} {\bibinfo {author} {\bibfnamefont {S.}~\bibnamefont
  {Hwan~Chun}}, \bibinfo {author} {\bibfnamefont {J.-W.}\ \bibnamefont {Kim}},
  \bibinfo {author} {\bibfnamefont {J.}~\bibnamefont {Kim}}, \bibinfo {author}
  {\bibfnamefont {H.}~\bibnamefont {Zheng}}, \bibinfo {author} {\bibfnamefont
  {C.}~\bibnamefont {Stoumpos}}, \bibinfo {author} {\bibfnamefont
  {C.}~\bibnamefont {Malliakas}}, \bibinfo {author} {\bibfnamefont
  {J.}~\bibnamefont {Mitchell}}, \bibinfo {author} {\bibfnamefont
  {K.}~\bibnamefont {Mehlawat}}, \bibinfo {author} {\bibfnamefont
  {Y.}~\bibnamefont {Singh}}, \bibinfo {author} {\bibfnamefont
  {Y.}~\bibnamefont {Choi}}, \bibinfo {author} {\bibfnamefont {T.}~\bibnamefont
  {Gog}}, \bibinfo {author} {\bibfnamefont {A.}~\bibnamefont {Al-Zein}},
  \bibinfo {author} {\bibfnamefont {M.}~\bibnamefont {Sala}}, \bibinfo {author}
  {\bibfnamefont {M.}~\bibnamefont {Krisch}}, \bibinfo {author} {\bibfnamefont
  {J.}~\bibnamefont {Chaloupka}}, \bibinfo {author} {\bibfnamefont
  {G.}~\bibnamefont {Jackeli}}, \bibinfo {author} {\bibfnamefont
  {G.}~\bibnamefont {Khaliullin}},\ and\ \bibinfo {author} {\bibfnamefont
  {B.~J.}\ \bibnamefont {Kim}},\ }\bibfield  {title} {\bibinfo {title} {Direct
  evidence for dominant bond-directional interactions in a honeycomb lattice
  iridate {$\text{Na}_2\text{IrO}_3$}},\ }\href
  {https://doi.org/10.1038/nphys3322} {\bibfield  {journal} {\bibinfo
  {journal} {Nature Physics}\ }\textbf {\bibinfo {volume} {11}},\ \bibinfo
  {pages} {462} (\bibinfo {year} {2015})}\BibitemShut {NoStop}%
\bibitem [{\citenamefont {Plumb}\ \emph {et~al.}(2014)\citenamefont {Plumb},
  \citenamefont {Clancy}, \citenamefont {Sandilands}, \citenamefont {Shankar},
  \citenamefont {Hu}, \citenamefont {Burch}, \citenamefont {Kee},\ and\
  \citenamefont {Kim}}]{plumb_alpha_RuCl3_2014}%
  \BibitemOpen
  \bibfield  {author} {\bibinfo {author} {\bibfnamefont {K.~W.}\ \bibnamefont
  {Plumb}}, \bibinfo {author} {\bibfnamefont {J.~P.}\ \bibnamefont {Clancy}},
  \bibinfo {author} {\bibfnamefont {L.~J.}\ \bibnamefont {Sandilands}},
  \bibinfo {author} {\bibfnamefont {V.~V.}\ \bibnamefont {Shankar}}, \bibinfo
  {author} {\bibfnamefont {Y.~F.}\ \bibnamefont {Hu}}, \bibinfo {author}
  {\bibfnamefont {K.~S.}\ \bibnamefont {Burch}}, \bibinfo {author}
  {\bibfnamefont {H.-Y.}\ \bibnamefont {Kee}},\ and\ \bibinfo {author}
  {\bibfnamefont {Y.-J.}\ \bibnamefont {Kim}},\ }\bibfield  {title} {\bibinfo
  {title} {{$\alpha-\text{RuCl}_3$}: {A} spin-orbit assisted {Mott} insulator
  on a honeycomb lattice},\ }\href {https://doi.org/10.1103/PhysRevB.90.041112}
  {\bibfield  {journal} {\bibinfo  {journal} {Phys. Rev. B}\ }\textbf {\bibinfo
  {volume} {90}},\ \bibinfo {pages} {041112} (\bibinfo {year}
  {2014})}\BibitemShut {NoStop}%
\bibitem [{\citenamefont {Kubota}\ \emph {et~al.}(2015)\citenamefont {Kubota},
  \citenamefont {Tanaka}, \citenamefont {Ono}, \citenamefont {Narumi},\ and\
  \citenamefont {Kindo}}]{kubota_successive_2015}%
  \BibitemOpen
  \bibfield  {author} {\bibinfo {author} {\bibfnamefont {Y.}~\bibnamefont
  {Kubota}}, \bibinfo {author} {\bibfnamefont {H.}~\bibnamefont {Tanaka}},
  \bibinfo {author} {\bibfnamefont {T.}~\bibnamefont {Ono}}, \bibinfo {author}
  {\bibfnamefont {Y.}~\bibnamefont {Narumi}},\ and\ \bibinfo {author}
  {\bibfnamefont {K.}~\bibnamefont {Kindo}},\ }\bibfield  {title} {\bibinfo
  {title} {Successive magnetic phase transitions in {$\alpha-\text{RuCl}_3$}:
  {XY}-like frustrated magnet on the honeycomb lattice},\ }\href
  {https://doi.org/10.1103/PhysRevB.91.094422} {\bibfield  {journal} {\bibinfo
  {journal} {Phys. Rev. B}\ }\textbf {\bibinfo {volume} {91}},\ \bibinfo
  {pages} {094422} (\bibinfo {year} {2015})}\BibitemShut {NoStop}%
\bibitem [{\citenamefont {Majumder}\ \emph {et~al.}(2015)\citenamefont
  {Majumder}, \citenamefont {Schmidt}, \citenamefont {Rosner}, \citenamefont
  {Tsirlin}, \citenamefont {Yasuoka},\ and\ \citenamefont
  {Baenitz}}]{majumder_anisotropic_2015}%
  \BibitemOpen
  \bibfield  {author} {\bibinfo {author} {\bibfnamefont {M.}~\bibnamefont
  {Majumder}}, \bibinfo {author} {\bibfnamefont {M.}~\bibnamefont {Schmidt}},
  \bibinfo {author} {\bibfnamefont {H.}~\bibnamefont {Rosner}}, \bibinfo
  {author} {\bibfnamefont {A.~A.}\ \bibnamefont {Tsirlin}}, \bibinfo {author}
  {\bibfnamefont {H.}~\bibnamefont {Yasuoka}},\ and\ \bibinfo {author}
  {\bibfnamefont {M.}~\bibnamefont {Baenitz}},\ }\bibfield  {title} {\bibinfo
  {title} {Anisotropic {$\text{Ru}^{3+}$} {$4d^5$} magnetism in the
  {$\alpha-\text{RuCl}_3$} honeycomb system: {Susceptibility}, specific heat,
  and zero-field {NMR}},\ }\href {https://doi.org/10.1103/PhysRevB.91.180401}
  {\bibfield  {journal} {\bibinfo  {journal} {Phys. Rev. B}\ }\textbf {\bibinfo
  {volume} {91}},\ \bibinfo {pages} {180401} (\bibinfo {year}
  {2015})}\BibitemShut {NoStop}%
\bibitem [{\citenamefont {Johnson}\ \emph {et~al.}(2015)\citenamefont
  {Johnson}, \citenamefont {Williams}, \citenamefont {Haghighirad},
  \citenamefont {Singleton}, \citenamefont {Zapf}, \citenamefont {Manuel},
  \citenamefont {Mazin}, \citenamefont {Li}, \citenamefont {Jeschke},
  \citenamefont {Valentí},\ and\ \citenamefont
  {Coldea}}]{johnson_monoclinic_2015}%
  \BibitemOpen
  \bibfield  {author} {\bibinfo {author} {\bibfnamefont {R.~D.}\ \bibnamefont
  {Johnson}}, \bibinfo {author} {\bibfnamefont {S.~C.}\ \bibnamefont
  {Williams}}, \bibinfo {author} {\bibfnamefont {A.~A.}\ \bibnamefont
  {Haghighirad}}, \bibinfo {author} {\bibfnamefont {J.}~\bibnamefont
  {Singleton}}, \bibinfo {author} {\bibfnamefont {V.}~\bibnamefont {Zapf}},
  \bibinfo {author} {\bibfnamefont {P.}~\bibnamefont {Manuel}}, \bibinfo
  {author} {\bibfnamefont {I.~I.}\ \bibnamefont {Mazin}}, \bibinfo {author}
  {\bibfnamefont {Y.}~\bibnamefont {Li}}, \bibinfo {author} {\bibfnamefont
  {H.~O.}\ \bibnamefont {Jeschke}}, \bibinfo {author} {\bibfnamefont
  {R.}~\bibnamefont {Valentí}},\ and\ \bibinfo {author} {\bibfnamefont
  {R.}~\bibnamefont {Coldea}},\ }\bibfield  {title} {\bibinfo {title}
  {Monoclinic crystal structure of {$\alpha-\text{RuCl}_3$} and the zigzag
  antiferromagnetic ground state},\ }\href
  {https://doi.org/10.1103/PhysRevB.92.235119} {\bibfield  {journal} {\bibinfo
  {journal} {Phys. Rev. B}\ }\textbf {\bibinfo {volume} {92}},\ \bibinfo
  {pages} {235119} (\bibinfo {year} {2015})}\BibitemShut {NoStop}%
\bibitem [{\citenamefont {Banerjee}\ \emph {et~al.}(2016)\citenamefont
  {Banerjee}, \citenamefont {Bridges}, \citenamefont {Yan}, \citenamefont
  {Aczel}, \citenamefont {Li}, \citenamefont {Stone}, \citenamefont {Granroth},
  \citenamefont {Lumsden}, \citenamefont {Yiu}, \citenamefont {Knolle},
  \citenamefont {Bhattacharjee}, \citenamefont {Kovrizhin}, \citenamefont
  {Moessner}, \citenamefont {Tennant}, \citenamefont {Mandrus},\ and\
  \citenamefont {Nagler}}]{banerjee_proximate_2016}%
  \BibitemOpen
  \bibfield  {author} {\bibinfo {author} {\bibfnamefont {A.}~\bibnamefont
  {Banerjee}}, \bibinfo {author} {\bibfnamefont {C.~A.}\ \bibnamefont
  {Bridges}}, \bibinfo {author} {\bibfnamefont {J.-Q.}\ \bibnamefont {Yan}},
  \bibinfo {author} {\bibfnamefont {A.~A.}\ \bibnamefont {Aczel}}, \bibinfo
  {author} {\bibfnamefont {L.}~\bibnamefont {Li}}, \bibinfo {author}
  {\bibfnamefont {M.~B.}\ \bibnamefont {Stone}}, \bibinfo {author}
  {\bibfnamefont {G.~E.}\ \bibnamefont {Granroth}}, \bibinfo {author}
  {\bibfnamefont {M.~D.}\ \bibnamefont {Lumsden}}, \bibinfo {author}
  {\bibfnamefont {Y.}~\bibnamefont {Yiu}}, \bibinfo {author} {\bibfnamefont
  {J.}~\bibnamefont {Knolle}}, \bibinfo {author} {\bibfnamefont
  {S.}~\bibnamefont {Bhattacharjee}}, \bibinfo {author} {\bibfnamefont {D.~L.}\
  \bibnamefont {Kovrizhin}}, \bibinfo {author} {\bibfnamefont {R.}~\bibnamefont
  {Moessner}}, \bibinfo {author} {\bibfnamefont {D.~A.}\ \bibnamefont
  {Tennant}}, \bibinfo {author} {\bibfnamefont {D.~G.}\ \bibnamefont
  {Mandrus}},\ and\ \bibinfo {author} {\bibfnamefont {S.~E.}\ \bibnamefont
  {Nagler}},\ }\bibfield  {title} {\bibinfo {title} {Proximate {Kitaev} quantum
  spin liquid behaviour in a honeycomb magnet},\ }\href
  {https://doi.org/10.1038/nmat4604} {\bibfield  {journal} {\bibinfo  {journal}
  {Nature Materials}\ }\textbf {\bibinfo {volume} {15}},\ \bibinfo {pages}
  {733} (\bibinfo {year} {2016})}\BibitemShut {NoStop}%
\bibitem [{\citenamefont {Wang}\ \emph {et~al.}(2017)\citenamefont {Wang},
  \citenamefont {Dong}, \citenamefont {Yu},\ and\ \citenamefont
  {Li}}]{wang_theoretical_2017}%
  \BibitemOpen
  \bibfield  {author} {\bibinfo {author} {\bibfnamefont {W.}~\bibnamefont
  {Wang}}, \bibinfo {author} {\bibfnamefont {Z.-Y.}\ \bibnamefont {Dong}},
  \bibinfo {author} {\bibfnamefont {S.-L.}\ \bibnamefont {Yu}},\ and\ \bibinfo
  {author} {\bibfnamefont {J.-X.}\ \bibnamefont {Li}},\ }\bibfield  {title}
  {\bibinfo {title} {Theoretical investigation of magnetic dynamics in
  {$\alpha-\text{RuCl}_3$}},\ }\href
  {https://doi.org/10.1103/PhysRevB.96.115103} {\bibfield  {journal} {\bibinfo
  {journal} {Phys. Rev. B}\ }\textbf {\bibinfo {volume} {96}},\ \bibinfo
  {pages} {115103} (\bibinfo {year} {2017})}\BibitemShut {NoStop}%
\bibitem [{\citenamefont {Ran}\ \emph {et~al.}(2017)\citenamefont {Ran},
  \citenamefont {Wang}, \citenamefont {Wang}, \citenamefont {Dong},
  \citenamefont {Ren}, \citenamefont {Bao}, \citenamefont {Li}, \citenamefont
  {Ma}, \citenamefont {Gan}, \citenamefont {Zhang}, \citenamefont {Park},
  \citenamefont {Deng}, \citenamefont {Danilkin}, \citenamefont {Yu},
  \citenamefont {Li},\ and\ \citenamefont
  {Wen}}]{SpinWaveExcitations_Ran_2017}%
  \BibitemOpen
  \bibfield  {author} {\bibinfo {author} {\bibfnamefont {K.}~\bibnamefont
  {Ran}}, \bibinfo {author} {\bibfnamefont {J.}~\bibnamefont {Wang}}, \bibinfo
  {author} {\bibfnamefont {W.}~\bibnamefont {Wang}}, \bibinfo {author}
  {\bibfnamefont {Z.-Y.}\ \bibnamefont {Dong}}, \bibinfo {author}
  {\bibfnamefont {X.}~\bibnamefont {Ren}}, \bibinfo {author} {\bibfnamefont
  {S.}~\bibnamefont {Bao}}, \bibinfo {author} {\bibfnamefont {S.}~\bibnamefont
  {Li}}, \bibinfo {author} {\bibfnamefont {Z.}~\bibnamefont {Ma}}, \bibinfo
  {author} {\bibfnamefont {Y.}~\bibnamefont {Gan}}, \bibinfo {author}
  {\bibfnamefont {Y.}~\bibnamefont {Zhang}}, \bibinfo {author} {\bibfnamefont
  {J.~T.}\ \bibnamefont {Park}}, \bibinfo {author} {\bibfnamefont
  {G.}~\bibnamefont {Deng}}, \bibinfo {author} {\bibfnamefont {S.}~\bibnamefont
  {Danilkin}}, \bibinfo {author} {\bibfnamefont {S.-L.}\ \bibnamefont {Yu}},
  \bibinfo {author} {\bibfnamefont {J.-X.}\ \bibnamefont {Li}},\ and\ \bibinfo
  {author} {\bibfnamefont {J.}~\bibnamefont {Wen}},\ }\bibfield  {title}
  {\bibinfo {title} {Spin-{{Wave Excitations Evidencing}} the {{Kitaev
  Interaction}} in {{Single Crystalline}} {$\alpha$} - {{RuCl}} 3},\ }\href
  {https://doi.org/10.1103/PhysRevLett.118.107203} {\bibfield  {journal}
  {\bibinfo  {journal} {Physical Review Letters}\ }\textbf {\bibinfo {volume}
  {118}},\ \bibinfo {pages} {107203} (\bibinfo {year} {2017})}\BibitemShut
  {NoStop}%
\bibitem [{\citenamefont {Liu}\ and\ \citenamefont
  {Khaliullin}(2018)}]{liu_pseudospin_2018}%
  \BibitemOpen
  \bibfield  {author} {\bibinfo {author} {\bibfnamefont {H.}~\bibnamefont
  {Liu}}\ and\ \bibinfo {author} {\bibfnamefont {G.}~\bibnamefont
  {Khaliullin}},\ }\bibfield  {title} {\bibinfo {title} {Pseudospin exchange
  interactions in {$d^7$} cobalt compounds: {Possible} realization of the
  {Kitaev} model},\ }\href {https://doi.org/10.1103/PhysRevB.97.014407}
  {\bibfield  {journal} {\bibinfo  {journal} {Phys. Rev. B}\ }\textbf {\bibinfo
  {volume} {97}},\ \bibinfo {pages} {014407} (\bibinfo {year}
  {2018})}\BibitemShut {NoStop}%
\bibitem [{\citenamefont {Sano}\ \emph {et~al.}(2018)\citenamefont {Sano},
  \citenamefont {Kato},\ and\ \citenamefont
  {Motome}}]{sano_kitaev-heisenberg_2018}%
  \BibitemOpen
  \bibfield  {author} {\bibinfo {author} {\bibfnamefont {R.}~\bibnamefont
  {Sano}}, \bibinfo {author} {\bibfnamefont {Y.}~\bibnamefont {Kato}},\ and\
  \bibinfo {author} {\bibfnamefont {Y.}~\bibnamefont {Motome}},\ }\bibfield
  {title} {\bibinfo {title} {Kitaev-{Heisenberg} {Hamiltonian} for high-spin
  {$d^7$} {Mott} insulators},\ }\href
  {https://doi.org/10.1103/PhysRevB.97.014408} {\bibfield  {journal} {\bibinfo
  {journal} {Phys. Rev. B}\ }\textbf {\bibinfo {volume} {97}},\ \bibinfo
  {pages} {014408} (\bibinfo {year} {2018})}\BibitemShut {NoStop}%
\bibitem [{\citenamefont {Liu}\ \emph {et~al.}(2020)\citenamefont {Liu},
  \citenamefont {Chaloupka},\ and\ \citenamefont
  {Khaliullin}}]{KitaevSpin_Liu_2020}%
  \BibitemOpen
  \bibfield  {author} {\bibinfo {author} {\bibfnamefont {H.}~\bibnamefont
  {Liu}}, \bibinfo {author} {\bibfnamefont {J.}~\bibnamefont {Chaloupka}},\
  and\ \bibinfo {author} {\bibfnamefont {G.}~\bibnamefont {Khaliullin}},\
  }\bibfield  {title} {\bibinfo {title} {Kitaev {{Spin Liquid}} in $3d$
  {{Transition Metal Compounds}}},\ }\href
  {https://doi.org/10.1103/PhysRevLett.125.047201} {\bibfield  {journal}
  {\bibinfo  {journal} {Physical Review Letters}\ }\textbf {\bibinfo {volume}
  {125}},\ \bibinfo {pages} {047201} (\bibinfo {year} {2020})}\BibitemShut
  {NoStop}%
\bibitem [{\citenamefont {Liu}\ and\ \citenamefont
  {Kee}(2023)}]{NonKitaevKitaev_Liu_2023}%
  \BibitemOpen
  \bibfield  {author} {\bibinfo {author} {\bibfnamefont {X.}~\bibnamefont
  {Liu}}\ and\ \bibinfo {author} {\bibfnamefont {H.-Y.}\ \bibnamefont {Kee}},\
  }\bibfield  {title} {\bibinfo {title} {Non-{{Kitaev}} versus {{Kitaev}}
  honeycomb cobaltates},\ }\href {https://doi.org/10.1103/PhysRevB.107.054420}
  {\bibfield  {journal} {\bibinfo  {journal} {Physical Review B}\ }\textbf
  {\bibinfo {volume} {107}},\ \bibinfo {pages} {054420} (\bibinfo {year}
  {2023})}\BibitemShut {NoStop}%
\bibitem [{\citenamefont {Chaloupka}\ and\ \citenamefont
  {Khaliullin}(2016)}]{MagneticAnisotropy_Chaloupka_2016}%
  \BibitemOpen
  \bibfield  {author} {\bibinfo {author} {\bibfnamefont {J.}~\bibnamefont
  {Chaloupka}}\ and\ \bibinfo {author} {\bibfnamefont {G.}~\bibnamefont
  {Khaliullin}},\ }\bibfield  {title} {\bibinfo {title} {Magnetic anisotropy in
  the {{Kitaev}} model systems {{$\text{Na}_2$}}{{$\text{IrO}_3$}} and
  {{$\text{RuCl}_3$}}},\ }\href {https://doi.org/10.1103/PhysRevB.94.064435}
  {\bibfield  {journal} {\bibinfo  {journal} {Physical Review B}\ }\textbf
  {\bibinfo {volume} {94}},\ \bibinfo {pages} {064435} (\bibinfo {year}
  {2016})}\BibitemShut {NoStop}%
\bibitem [{\citenamefont {Winter}\ \emph {et~al.}(2017)\citenamefont {Winter},
  \citenamefont {Tsirlin}, \citenamefont {Daghofer}, \citenamefont {Van
  Den~Brink}, \citenamefont {Singh}, \citenamefont {Gegenwart},\ and\
  \citenamefont {Valent{\'i}}}]{ModelsMaterials_Winter_2017}%
  \BibitemOpen
  \bibfield  {author} {\bibinfo {author} {\bibfnamefont {S.~M.}\ \bibnamefont
  {Winter}}, \bibinfo {author} {\bibfnamefont {A.~A.}\ \bibnamefont {Tsirlin}},
  \bibinfo {author} {\bibfnamefont {M.}~\bibnamefont {Daghofer}}, \bibinfo
  {author} {\bibfnamefont {J.}~\bibnamefont {Van Den~Brink}}, \bibinfo {author}
  {\bibfnamefont {Y.}~\bibnamefont {Singh}}, \bibinfo {author} {\bibfnamefont
  {P.}~\bibnamefont {Gegenwart}},\ and\ \bibinfo {author} {\bibfnamefont
  {R.}~\bibnamefont {Valent{\'i}}},\ }\bibfield  {title} {\bibinfo {title}
  {Models and materials for generalized {{Kitaev}} magnetism},\ }\href
  {https://doi.org/10.1088/1361-648X/aa8cf5} {\bibfield  {journal} {\bibinfo
  {journal} {Journal of Physics: Condensed Matter}\ }\textbf {\bibinfo {volume}
  {29}},\ \bibinfo {pages} {493002} (\bibinfo {year} {2017})}\BibitemShut
  {NoStop}%
\bibitem [{\citenamefont {Takagi}\ \emph {et~al.}(2019)\citenamefont {Takagi},
  \citenamefont {Takayama}, \citenamefont {Jackeli}, \citenamefont
  {Khaliullin},\ and\ \citenamefont {Nagler}}]{ConceptRealization_Takagi_2019}%
  \BibitemOpen
  \bibfield  {author} {\bibinfo {author} {\bibfnamefont {H.}~\bibnamefont
  {Takagi}}, \bibinfo {author} {\bibfnamefont {T.}~\bibnamefont {Takayama}},
  \bibinfo {author} {\bibfnamefont {G.}~\bibnamefont {Jackeli}}, \bibinfo
  {author} {\bibfnamefont {G.}~\bibnamefont {Khaliullin}},\ and\ \bibinfo
  {author} {\bibfnamefont {S.~E.}\ \bibnamefont {Nagler}},\ }\bibfield  {title}
  {\bibinfo {title} {Concept and realization of {{Kitaev}} quantum spin
  liquids},\ }\href {https://doi.org/10.1038/s42254-019-0038-2} {\bibfield
  {journal} {\bibinfo  {journal} {Nature Reviews Physics}\ }\textbf {\bibinfo
  {volume} {1}},\ \bibinfo {pages} {264} (\bibinfo {year} {2019})}\BibitemShut
  {NoStop}%
\bibitem [{\citenamefont {Motome}\ and\ \citenamefont
  {Nasu}(2020)}]{HuntingMajorana_Motome_2020}%
  \BibitemOpen
  \bibfield  {author} {\bibinfo {author} {\bibfnamefont {Y.}~\bibnamefont
  {Motome}}\ and\ \bibinfo {author} {\bibfnamefont {J.}~\bibnamefont {Nasu}},\
  }\bibfield  {title} {\bibinfo {title} {Hunting {{Majorana Fermions}} in
  {{Kitaev Magnets}}},\ }\href {https://doi.org/10.7566/JPSJ.89.012002}
  {\bibfield  {journal} {\bibinfo  {journal} {Journal of the Physical Society
  of Japan}\ }\textbf {\bibinfo {volume} {89}},\ \bibinfo {pages} {012002}
  (\bibinfo {year} {2020})}\BibitemShut {NoStop}%
\bibitem [{\citenamefont {Trebst}\ and\ \citenamefont
  {Hickey}(2022)}]{KitaevMaterials_Trebst_2022}%
  \BibitemOpen
  \bibfield  {author} {\bibinfo {author} {\bibfnamefont {S.}~\bibnamefont
  {Trebst}}\ and\ \bibinfo {author} {\bibfnamefont {C.}~\bibnamefont
  {Hickey}},\ }\bibfield  {title} {\bibinfo {title} {Kitaev materials},\ }\href
  {https://doi.org/10.1016/j.physrep.2021.11.003} {\bibfield  {journal}
  {\bibinfo  {journal} {Physics Reports}\ }\textbf {\bibinfo {volume} {950}},\
  \bibinfo {pages} {1} (\bibinfo {year} {2022})}\BibitemShut {NoStop}%
\bibitem [{\citenamefont {Liu}\ \emph {et~al.}(2022)\citenamefont {Liu},
  \citenamefont {Chaloupka},\ and\ \citenamefont
  {Khaliullin}}]{ExchangeInteractions_Liu_2022}%
  \BibitemOpen
  \bibfield  {author} {\bibinfo {author} {\bibfnamefont {H.}~\bibnamefont
  {Liu}}, \bibinfo {author} {\bibfnamefont {J.}~\bibnamefont {Chaloupka}},\
  and\ \bibinfo {author} {\bibfnamefont {G.}~\bibnamefont {Khaliullin}},\
  }\bibfield  {title} {\bibinfo {title} {Exchange interactions in {$d^5$}
  kitaev materials: From {$\text{Na}_2$}{$\text{IrO}_3$} to {$\alpha$} -
  {$\text{RuCl}_3$}},\ }\href {https://doi.org/10.1103/PhysRevB.105.214411}
  {\bibfield  {journal} {\bibinfo  {journal} {Physical Review B}\ }\textbf
  {\bibinfo {volume} {105}},\ \bibinfo {pages} {214411} (\bibinfo {year}
  {2022})}\BibitemShut {NoStop}%
\bibitem [{\citenamefont {Rousochatzakis}\ \emph {et~al.}(2024)\citenamefont
  {Rousochatzakis}, \citenamefont {Perkins}, \citenamefont {Luo},\ and\
  \citenamefont {Kee}}]{KitaevPhysics_Rousochatzakis_2024}%
  \BibitemOpen
  \bibfield  {author} {\bibinfo {author} {\bibfnamefont {I.}~\bibnamefont
  {Rousochatzakis}}, \bibinfo {author} {\bibfnamefont {N.~B.}\ \bibnamefont
  {Perkins}}, \bibinfo {author} {\bibfnamefont {Q.}~\bibnamefont {Luo}},\ and\
  \bibinfo {author} {\bibfnamefont {H.-Y.}\ \bibnamefont {Kee}},\ }\bibfield
  {title} {\bibinfo {title} {Beyond {{Kitaev}} physics in strong spin-orbit
  coupled magnets},\ }\href {https://doi.org/10.1088/1361-6633/ad208d}
  {\bibfield  {journal} {\bibinfo  {journal} {Reports on Progress in Physics}\
  }\textbf {\bibinfo {volume} {87}},\ \bibinfo {pages} {026502} (\bibinfo
  {year} {2024})}\BibitemShut {NoStop}%
\bibitem [{\citenamefont {Lado}\ and\ \citenamefont
  {Fernández-Rossier}(2017)}]{lado_origin_2017}%
  \BibitemOpen
  \bibfield  {author} {\bibinfo {author} {\bibfnamefont {J.~L.}\ \bibnamefont
  {Lado}}\ and\ \bibinfo {author} {\bibfnamefont {J.}~\bibnamefont
  {Fernández-Rossier}},\ }\bibfield  {title} {\bibinfo {title} {On the origin
  of magnetic anisotropy in two dimensional $\text{CrI}_3$},\ }\href
  {https://doi.org/10.1088/2053-1583/aa75ed} {\bibfield  {journal} {\bibinfo
  {journal} {2D Materials}\ }\textbf {\bibinfo {volume} {4}},\ \bibinfo {pages}
  {035002} (\bibinfo {year} {2017})}\BibitemShut {NoStop}%
\bibitem [{\citenamefont {Xu}\ \emph {et~al.}(2018)\citenamefont {Xu},
  \citenamefont {Feng}, \citenamefont {Xiang},\ and\ \citenamefont
  {Bellaiche}}]{xu_interplay_2018}%
  \BibitemOpen
  \bibfield  {author} {\bibinfo {author} {\bibfnamefont {C.}~\bibnamefont
  {Xu}}, \bibinfo {author} {\bibfnamefont {J.}~\bibnamefont {Feng}}, \bibinfo
  {author} {\bibfnamefont {H.}~\bibnamefont {Xiang}},\ and\ \bibinfo {author}
  {\bibfnamefont {L.}~\bibnamefont {Bellaiche}},\ }\bibfield  {title} {\bibinfo
  {title} {Interplay between {Kitaev} interaction and single ion anisotropy in
  ferromagnetic $\text{CrI}_3$ and $\text{CrGeTe}_3$ monolayers},\ }\href
  {https://doi.org/10.1038/s41524-018-0115-6} {\bibfield  {journal} {\bibinfo
  {journal} {npj Computational Materials}\ }\textbf {\bibinfo {volume} {4}},\
  \bibinfo {pages} {57} (\bibinfo {year} {2018})}\BibitemShut {NoStop}%
\bibitem [{\citenamefont {Stavropoulos}\ \emph {et~al.}(2019)\citenamefont
  {Stavropoulos}, \citenamefont {Pereira},\ and\ \citenamefont
  {Kee}}]{stavropoulos_microscopic_2019}%
  \BibitemOpen
  \bibfield  {author} {\bibinfo {author} {\bibfnamefont {P.~P.}\ \bibnamefont
  {Stavropoulos}}, \bibinfo {author} {\bibfnamefont {D.}~\bibnamefont
  {Pereira}},\ and\ \bibinfo {author} {\bibfnamefont {H.-Y.}\ \bibnamefont
  {Kee}},\ }\bibfield  {title} {\bibinfo {title} {Microscopic {Mechanism} for a
  {Higher}-{Spin} {Kitaev} {Model}},\ }\href
  {https://doi.org/10.1103/PhysRevLett.123.037203} {\bibfield  {journal}
  {\bibinfo  {journal} {Phys. Rev. Lett.}\ }\textbf {\bibinfo {volume} {123}},\
  \bibinfo {pages} {037203} (\bibinfo {year} {2019})}\BibitemShut {NoStop}%
\bibitem [{\citenamefont {Kim}\ \emph {et~al.}(2019)\citenamefont {Kim},
  \citenamefont {Kim}, \citenamefont {Ko}, \citenamefont {Seo}, \citenamefont
  {Kim}, \citenamefont {Jang}, \citenamefont {Kim}, \citenamefont {Kim},
  \citenamefont {Cheong},\ and\ \citenamefont {Park}}]{kim_giant_2019}%
  \BibitemOpen
  \bibfield  {author} {\bibinfo {author} {\bibfnamefont {D.-H.}\ \bibnamefont
  {Kim}}, \bibinfo {author} {\bibfnamefont {K.}~\bibnamefont {Kim}}, \bibinfo
  {author} {\bibfnamefont {K.-T.}\ \bibnamefont {Ko}}, \bibinfo {author}
  {\bibfnamefont {J.}~\bibnamefont {Seo}}, \bibinfo {author} {\bibfnamefont
  {J.~S.}\ \bibnamefont {Kim}}, \bibinfo {author} {\bibfnamefont {T.-H.}\
  \bibnamefont {Jang}}, \bibinfo {author} {\bibfnamefont {Y.}~\bibnamefont
  {Kim}}, \bibinfo {author} {\bibfnamefont {J.-Y.}\ \bibnamefont {Kim}},
  \bibinfo {author} {\bibfnamefont {S.-W.}\ \bibnamefont {Cheong}},\ and\
  \bibinfo {author} {\bibfnamefont {J.-H.}\ \bibnamefont {Park}},\ }\bibfield
  {title} {\bibinfo {title} {Giant {Magnetic} {Anisotropy} {Induced} by
  {Ligand} {LS} {Coupling} in {Layered} {Cr} {Compounds}},\ }\href
  {https://doi.org/10.1103/PhysRevLett.122.207201} {\bibfield  {journal}
  {\bibinfo  {journal} {Phys. Rev. Lett.}\ }\textbf {\bibinfo {volume} {122}},\
  \bibinfo {pages} {207201} (\bibinfo {year} {2019})}\BibitemShut {NoStop}%
\bibitem [{\citenamefont {Lee}\ \emph {et~al.}(2020)\citenamefont {Lee},
  \citenamefont {Utermohlen}, \citenamefont {Weber}, \citenamefont {Hwang},
  \citenamefont {Zhang}, \citenamefont {van Tol}, \citenamefont {Goldberger},
  \citenamefont {Trivedi},\ and\ \citenamefont
  {Hammel}}]{lee_fundamental_2020}%
  \BibitemOpen
  \bibfield  {author} {\bibinfo {author} {\bibfnamefont {I.}~\bibnamefont
  {Lee}}, \bibinfo {author} {\bibfnamefont {F.~G.}\ \bibnamefont {Utermohlen}},
  \bibinfo {author} {\bibfnamefont {D.}~\bibnamefont {Weber}}, \bibinfo
  {author} {\bibfnamefont {K.}~\bibnamefont {Hwang}}, \bibinfo {author}
  {\bibfnamefont {C.}~\bibnamefont {Zhang}}, \bibinfo {author} {\bibfnamefont
  {J.}~\bibnamefont {van Tol}}, \bibinfo {author} {\bibfnamefont {J.~E.}\
  \bibnamefont {Goldberger}}, \bibinfo {author} {\bibfnamefont
  {N.}~\bibnamefont {Trivedi}},\ and\ \bibinfo {author} {\bibfnamefont {P.~C.}\
  \bibnamefont {Hammel}},\ }\bibfield  {title} {\bibinfo {title} {Fundamental
  {Spin} {Interactions} {Underlying} the {Magnetic} {Anisotropy} in the
  {Kitaev} {Ferromagnet} {$\text{CrI}_3$}},\ }\href
  {https://doi.org/10.1103/PhysRevLett.124.017201} {\bibfield  {journal}
  {\bibinfo  {journal} {Phys. Rev. Lett.}\ }\textbf {\bibinfo {volume} {124}},\
  \bibinfo {pages} {017201} (\bibinfo {year} {2020})}\BibitemShut {NoStop}%
\bibitem [{\citenamefont {Xu}\ \emph {et~al.}(2020)\citenamefont {Xu},
  \citenamefont {Feng}, \citenamefont {Kawamura}, \citenamefont {Yamaji},
  \citenamefont {Nahas}, \citenamefont {Prokhorenko}, \citenamefont {Qi},
  \citenamefont {Xiang},\ and\ \citenamefont
  {Bellaiche}}]{PossibleKitaev_Xu_2020}%
  \BibitemOpen
  \bibfield  {author} {\bibinfo {author} {\bibfnamefont {C.}~\bibnamefont
  {Xu}}, \bibinfo {author} {\bibfnamefont {J.}~\bibnamefont {Feng}}, \bibinfo
  {author} {\bibfnamefont {M.}~\bibnamefont {Kawamura}}, \bibinfo {author}
  {\bibfnamefont {Y.}~\bibnamefont {Yamaji}}, \bibinfo {author} {\bibfnamefont
  {Y.}~\bibnamefont {Nahas}}, \bibinfo {author} {\bibfnamefont
  {S.}~\bibnamefont {Prokhorenko}}, \bibinfo {author} {\bibfnamefont
  {Y.}~\bibnamefont {Qi}}, \bibinfo {author} {\bibfnamefont {H.}~\bibnamefont
  {Xiang}},\ and\ \bibinfo {author} {\bibfnamefont {L.}~\bibnamefont
  {Bellaiche}},\ }\bibfield  {title} {\bibinfo {title} {Possible {{Kitaev
  Quantum Spin Liquid State}} in {{2D Materials}} with {{S}} = 3 / 2},\ }\href
  {https://doi.org/10.1103/PhysRevLett.124.087205} {\bibfield  {journal}
  {\bibinfo  {journal} {Physical Review Letters}\ }\textbf {\bibinfo {volume}
  {124}},\ \bibinfo {pages} {087205} (\bibinfo {year} {2020})}\BibitemShut
  {NoStop}%
\bibitem [{\citenamefont {Stavropoulos}\ \emph {et~al.}(2021)\citenamefont
  {Stavropoulos}, \citenamefont {Liu},\ and\ \citenamefont
  {Kee}}]{stavropoulos_magnetic_2021}%
  \BibitemOpen
  \bibfield  {author} {\bibinfo {author} {\bibfnamefont {P.~P.}\ \bibnamefont
  {Stavropoulos}}, \bibinfo {author} {\bibfnamefont {X.}~\bibnamefont {Liu}},\
  and\ \bibinfo {author} {\bibfnamefont {H.-Y.}\ \bibnamefont {Kee}},\
  }\bibfield  {title} {\bibinfo {title} {Magnetic anisotropy in spin-3/2 with
  heavy ligand in honeycomb {Mott} insulators: {Application} to
  {$\text{CrI}_3$}},\ }\href {https://doi.org/10.1103/PhysRevResearch.3.013216}
  {\bibfield  {journal} {\bibinfo  {journal} {Phys. Rev. Res.}\ }\textbf
  {\bibinfo {volume} {3}},\ \bibinfo {pages} {013216} (\bibinfo {year}
  {2021})}\BibitemShut {NoStop}%
\bibitem [{\citenamefont {Bandyopadhyay}\ \emph {et~al.}(2022)\citenamefont
  {Bandyopadhyay}, \citenamefont {Buessen}, \citenamefont {Das}, \citenamefont
  {Utermohlen}, \citenamefont {Trivedi}, \citenamefont {Paramekanti},\ and\
  \citenamefont {Dasgupta}}]{bandyopadhyay_exchange_2022}%
  \BibitemOpen
  \bibfield  {author} {\bibinfo {author} {\bibfnamefont {S.}~\bibnamefont
  {Bandyopadhyay}}, \bibinfo {author} {\bibfnamefont {F.~L.}\ \bibnamefont
  {Buessen}}, \bibinfo {author} {\bibfnamefont {R.}~\bibnamefont {Das}},
  \bibinfo {author} {\bibfnamefont {F.~G.}\ \bibnamefont {Utermohlen}},
  \bibinfo {author} {\bibfnamefont {N.}~\bibnamefont {Trivedi}}, \bibinfo
  {author} {\bibfnamefont {A.}~\bibnamefont {Paramekanti}},\ and\ \bibinfo
  {author} {\bibfnamefont {I.}~\bibnamefont {Dasgupta}},\ }\bibfield  {title}
  {\bibinfo {title} {Exchange interactions and spin dynamics in the layered
  honeycomb ferromagnet {$\text{CrI}_3$}},\ }\href
  {https://doi.org/10.1103/PhysRevB.105.184430} {\bibfield  {journal} {\bibinfo
   {journal} {Phys. Rev. B}\ }\textbf {\bibinfo {volume} {105}},\ \bibinfo
  {pages} {184430} (\bibinfo {year} {2022})}\BibitemShut {NoStop}%
\bibitem [{\citenamefont {Riedl}\ \emph {et~al.}(2022)\citenamefont {Riedl},
  \citenamefont {Amoroso}, \citenamefont {Backes}, \citenamefont {Razpopov},
  \citenamefont {Nguyen}, \citenamefont {Yamauchi}, \citenamefont {Barone},
  \citenamefont {Winter}, \citenamefont {Picozzi},\ and\ \citenamefont
  {Valentí}}]{riedl_microscopic_2022}%
  \BibitemOpen
  \bibfield  {author} {\bibinfo {author} {\bibfnamefont {K.}~\bibnamefont
  {Riedl}}, \bibinfo {author} {\bibfnamefont {D.}~\bibnamefont {Amoroso}},
  \bibinfo {author} {\bibfnamefont {S.}~\bibnamefont {Backes}}, \bibinfo
  {author} {\bibfnamefont {A.}~\bibnamefont {Razpopov}}, \bibinfo {author}
  {\bibfnamefont {T.~P.~T.}\ \bibnamefont {Nguyen}}, \bibinfo {author}
  {\bibfnamefont {K.}~\bibnamefont {Yamauchi}}, \bibinfo {author}
  {\bibfnamefont {P.}~\bibnamefont {Barone}}, \bibinfo {author} {\bibfnamefont
  {S.~M.}\ \bibnamefont {Winter}}, \bibinfo {author} {\bibfnamefont
  {S.}~\bibnamefont {Picozzi}},\ and\ \bibinfo {author} {\bibfnamefont
  {R.}~\bibnamefont {Valentí}},\ }\bibfield  {title} {\bibinfo {title}
  {Microscopic origin of magnetism in monolayer {$3d$} transition metal
  dihalides},\ }\href {https://doi.org/10.1103/PhysRevB.106.035156} {\bibfield
  {journal} {\bibinfo  {journal} {Phys. Rev. B}\ }\textbf {\bibinfo {volume}
  {106}},\ \bibinfo {pages} {035156} (\bibinfo {year} {2022})}\BibitemShut
  {NoStop}%
\bibitem [{\citenamefont {Shangguan}\ \emph {et~al.}(2023)\citenamefont
  {Shangguan}, \citenamefont {Bao}, \citenamefont {Dong}, \citenamefont {Xi},
  \citenamefont {Gao}, \citenamefont {Ma}, \citenamefont {Wang}, \citenamefont
  {Qi}, \citenamefont {Zhang}, \citenamefont {Huang}, \citenamefont {Liao},
  \citenamefont {Zhao}, \citenamefont {Zhang}, \citenamefont {Cheng},
  \citenamefont {Xu}, \citenamefont {Yu}, \citenamefont {Mole}, \citenamefont
  {Murai}, \citenamefont {{Ohira-Kawamura}}, \citenamefont {He}, \citenamefont
  {Hao}, \citenamefont {Yan}, \citenamefont {Song}, \citenamefont {Li},
  \citenamefont {Yu}, \citenamefont {Li},\ and\ \citenamefont
  {Wen}}]{OnethirdMagnetization_Shangguan_2023}%
  \BibitemOpen
  \bibfield  {author} {\bibinfo {author} {\bibfnamefont {Y.}~\bibnamefont
  {Shangguan}}, \bibinfo {author} {\bibfnamefont {S.}~\bibnamefont {Bao}},
  \bibinfo {author} {\bibfnamefont {Z.-Y.}\ \bibnamefont {Dong}}, \bibinfo
  {author} {\bibfnamefont {N.}~\bibnamefont {Xi}}, \bibinfo {author}
  {\bibfnamefont {Y.-P.}\ \bibnamefont {Gao}}, \bibinfo {author} {\bibfnamefont
  {Z.}~\bibnamefont {Ma}}, \bibinfo {author} {\bibfnamefont {W.}~\bibnamefont
  {Wang}}, \bibinfo {author} {\bibfnamefont {Z.}~\bibnamefont {Qi}}, \bibinfo
  {author} {\bibfnamefont {S.}~\bibnamefont {Zhang}}, \bibinfo {author}
  {\bibfnamefont {Z.}~\bibnamefont {Huang}}, \bibinfo {author} {\bibfnamefont
  {J.}~\bibnamefont {Liao}}, \bibinfo {author} {\bibfnamefont {X.}~\bibnamefont
  {Zhao}}, \bibinfo {author} {\bibfnamefont {B.}~\bibnamefont {Zhang}},
  \bibinfo {author} {\bibfnamefont {S.}~\bibnamefont {Cheng}}, \bibinfo
  {author} {\bibfnamefont {H.}~\bibnamefont {Xu}}, \bibinfo {author}
  {\bibfnamefont {D.}~\bibnamefont {Yu}}, \bibinfo {author} {\bibfnamefont
  {R.~A.}\ \bibnamefont {Mole}}, \bibinfo {author} {\bibfnamefont
  {N.}~\bibnamefont {Murai}}, \bibinfo {author} {\bibfnamefont
  {S.}~\bibnamefont {{Ohira-Kawamura}}}, \bibinfo {author} {\bibfnamefont
  {L.}~\bibnamefont {He}}, \bibinfo {author} {\bibfnamefont {J.}~\bibnamefont
  {Hao}}, \bibinfo {author} {\bibfnamefont {Q.-B.}\ \bibnamefont {Yan}},
  \bibinfo {author} {\bibfnamefont {F.}~\bibnamefont {Song}}, \bibinfo {author}
  {\bibfnamefont {W.}~\bibnamefont {Li}}, \bibinfo {author} {\bibfnamefont
  {S.-L.}\ \bibnamefont {Yu}}, \bibinfo {author} {\bibfnamefont {J.-X.}\
  \bibnamefont {Li}},\ and\ \bibinfo {author} {\bibfnamefont {J.}~\bibnamefont
  {Wen}},\ }\bibfield  {title} {\bibinfo {title} {A one-third magnetization
  plateau phase as evidence for the {{Kitaev}} interaction in a
  honeycomb-lattice antiferromagnet},\ }\href
  {https://doi.org/10.1038/s41567-023-02212-2} {\bibfield  {journal} {\bibinfo
  {journal} {Nature Physics}\ }\textbf {\bibinfo {volume} {19}},\ \bibinfo
  {pages} {1883} (\bibinfo {year} {2023})}\BibitemShut {NoStop}%
\bibitem [{\citenamefont {Luo}\ \emph {et~al.}(2024)\citenamefont {Luo},
  \citenamefont {Zhao}, \citenamefont {Li},\ and\ \citenamefont
  {Wang}}]{ChiralSpin_Luo_2024}%
  \BibitemOpen
  \bibfield  {author} {\bibinfo {author} {\bibfnamefont {Q.}~\bibnamefont
  {Luo}}, \bibinfo {author} {\bibfnamefont {J.}~\bibnamefont {Zhao}}, \bibinfo
  {author} {\bibfnamefont {J.}~\bibnamefont {Li}},\ and\ \bibinfo {author}
  {\bibfnamefont {X.}~\bibnamefont {Wang}},\ }\bibfield  {title} {\bibinfo
  {title} {Chiral spin state and nematic ferromagnet in the spin-1 {{Kitaev-
  $\Gamma$}} model},\ }\href {https://doi.org/10.1103/PhysRevB.110.035121}
  {\bibfield  {journal} {\bibinfo  {journal} {Physical Review B}\ }\textbf
  {\bibinfo {volume} {110}},\ \bibinfo {pages} {035121} (\bibinfo {year}
  {2024})}\BibitemShut {NoStop}%
\bibitem [{\citenamefont {Li}\ \emph {et~al.}(2017)\citenamefont {Li},
  \citenamefont {Li}, \citenamefont {Yu}, \citenamefont {Paramekanti},\ and\
  \citenamefont {Chen}}]{li_kitaev_2017}%
  \BibitemOpen
  \bibfield  {author} {\bibinfo {author} {\bibfnamefont {F.-Y.}\ \bibnamefont
  {Li}}, \bibinfo {author} {\bibfnamefont {Y.-D.}\ \bibnamefont {Li}}, \bibinfo
  {author} {\bibfnamefont {Y.}~\bibnamefont {Yu}}, \bibinfo {author}
  {\bibfnamefont {A.}~\bibnamefont {Paramekanti}},\ and\ \bibinfo {author}
  {\bibfnamefont {G.}~\bibnamefont {Chen}},\ }\bibfield  {title} {\bibinfo
  {title} {Kitaev materials beyond iridates: {Order} by quantum disorder and
  {Weyl} magnons in rare-earth double perovskites},\ }\href
  {https://doi.org/10.1103/PhysRevB.95.085132} {\bibfield  {journal} {\bibinfo
  {journal} {Physical Review B}\ }\textbf {\bibinfo {volume} {95}},\ \bibinfo
  {pages} {085132} (\bibinfo {year} {2017})}\BibitemShut {NoStop}%
\bibitem [{\citenamefont {Rau}\ and\ \citenamefont
  {Gingras}(2018)}]{rau_frustration_2018}%
  \BibitemOpen
  \bibfield  {author} {\bibinfo {author} {\bibfnamefont {J.~G.}\ \bibnamefont
  {Rau}}\ and\ \bibinfo {author} {\bibfnamefont {M.~J.~P.}\ \bibnamefont
  {Gingras}},\ }\bibfield  {title} {\bibinfo {title} {Frustration and
  anisotropic exchange in ytterbium magnets with edge-shared octahedra},\
  }\href {https://doi.org/10.1103/PhysRevB.98.054408} {\bibfield  {journal}
  {\bibinfo  {journal} {Physical Review B}\ }\textbf {\bibinfo {volume} {98}},\
  \bibinfo {pages} {054408} (\bibinfo {year} {2018})}\BibitemShut {NoStop}%
\bibitem [{\citenamefont {Motome}\ \emph {et~al.}(2020)\citenamefont {Motome},
  \citenamefont {Sano}, \citenamefont {Jang}, \citenamefont {Sugita},\ and\
  \citenamefont {Kato}}]{MaterialsDesign_Motome_2020}%
  \BibitemOpen
  \bibfield  {author} {\bibinfo {author} {\bibfnamefont {Y.}~\bibnamefont
  {Motome}}, \bibinfo {author} {\bibfnamefont {R.}~\bibnamefont {Sano}},
  \bibinfo {author} {\bibfnamefont {S.}~\bibnamefont {Jang}}, \bibinfo {author}
  {\bibfnamefont {Y.}~\bibnamefont {Sugita}},\ and\ \bibinfo {author}
  {\bibfnamefont {Y.}~\bibnamefont {Kato}},\ }\bibfield  {title} {\bibinfo
  {title} {Materials design of {{Kitaev}} spin liquids beyond the
  {{Jackeli}}--{{Khaliullin}} mechanism},\ }\href
  {https://doi.org/10.1088/1361-648X/ab8525} {\bibfield  {journal} {\bibinfo
  {journal} {Journal of Physics: Condensed Matter}\ }\textbf {\bibinfo {volume}
  {32}},\ \bibinfo {pages} {404001} (\bibinfo {year} {2020})}\BibitemShut
  {NoStop}%
\bibitem [{\citenamefont {Luo}\ and\ \citenamefont
  {Chen}(2020)}]{HoneycombRareearth_Luo_2020}%
  \BibitemOpen
  \bibfield  {author} {\bibinfo {author} {\bibfnamefont {Z.-X.}\ \bibnamefont
  {Luo}}\ and\ \bibinfo {author} {\bibfnamefont {G.}~\bibnamefont {Chen}},\
  }\bibfield  {title} {\bibinfo {title} {Honeycomb rare-earth magnets with
  anisotropic exchange interactions},\ }\href
  {https://doi.org/10.21468/SciPostPhysCore.3.1.004} {\bibfield  {journal}
  {\bibinfo  {journal} {SciPost Physics Core}\ }\textbf {\bibinfo {volume}
  {3}},\ \bibinfo {pages} {004} (\bibinfo {year} {2020})}\BibitemShut {NoStop}%
\bibitem [{\citenamefont {Jang}\ \emph {et~al.}(2019)\citenamefont {Jang},
  \citenamefont {Sano}, \citenamefont {Kato},\ and\ \citenamefont
  {Motome}}]{jang_antiferromagnetic_2019}%
  \BibitemOpen
  \bibfield  {author} {\bibinfo {author} {\bibfnamefont {S.-H.}\ \bibnamefont
  {Jang}}, \bibinfo {author} {\bibfnamefont {R.}~\bibnamefont {Sano}}, \bibinfo
  {author} {\bibfnamefont {Y.}~\bibnamefont {Kato}},\ and\ \bibinfo {author}
  {\bibfnamefont {Y.}~\bibnamefont {Motome}},\ }\bibfield  {title} {\bibinfo
  {title} {Antiferromagnetic {Kitaev} interaction in {$f$}-electron based
  honeycomb magnets},\ }\href {https://doi.org/10.1103/PhysRevB.99.241106}
  {\bibfield  {journal} {\bibinfo  {journal} {Phys. Rev. B}\ }\textbf {\bibinfo
  {volume} {99}},\ \bibinfo {pages} {241106} (\bibinfo {year}
  {2019})}\BibitemShut {NoStop}%
\bibitem [{\citenamefont {Jang}\ \emph {et~al.}(2020)\citenamefont {Jang},
  \citenamefont {Sano}, \citenamefont {Kato},\ and\ \citenamefont
  {Motome}}]{jang_computational_2020}%
  \BibitemOpen
  \bibfield  {author} {\bibinfo {author} {\bibfnamefont {S.-H.}\ \bibnamefont
  {Jang}}, \bibinfo {author} {\bibfnamefont {R.}~\bibnamefont {Sano}}, \bibinfo
  {author} {\bibfnamefont {Y.}~\bibnamefont {Kato}},\ and\ \bibinfo {author}
  {\bibfnamefont {Y.}~\bibnamefont {Motome}},\ }\bibfield  {title} {\bibinfo
  {title} {Computational design of {$f$}-electron {Kitaev} magnets: {Honeycomb}
  and hyperhoneycomb compounds {$\text{A}_2\text{PrO}_3$} ({\text{a}=} alkali
  metals)},\ }\href {https://doi.org/10.1103/PhysRevMaterials.4.104420}
  {\bibfield  {journal} {\bibinfo  {journal} {Phys. Rev. Mater.}\ }\textbf
  {\bibinfo {volume} {4}},\ \bibinfo {pages} {104420} (\bibinfo {year}
  {2020})}\BibitemShut {NoStop}%
\bibitem [{\citenamefont {Jang}\ and\ \citenamefont
  {Motome}(2024{\natexlab{a}})}]{jang_exchange_2024}%
  \BibitemOpen
  \bibfield  {author} {\bibinfo {author} {\bibfnamefont {S.-H.}\ \bibnamefont
  {Jang}}\ and\ \bibinfo {author} {\bibfnamefont {Y.}~\bibnamefont {Motome}},\
  }\bibfield  {title} {\bibinfo {title} {Exchange interactions in the
  rare-earth magnets {$\text{A}_2\text{PrO}_3$} ({$\text{A}=$}alkali metals)},\
  }\href {https://doi.org/10.1103/PhysRevB.110.155124} {\bibfield  {journal}
  {\bibinfo  {journal} {Phys. Rev. B}\ }\textbf {\bibinfo {volume} {110}},\
  \bibinfo {pages} {155124} (\bibinfo {year} {2024}{\natexlab{a}})}\BibitemShut
  {NoStop}%
\bibitem [{\citenamefont {Banerjee}\ \emph {et~al.}(2018)\citenamefont
  {Banerjee}, \citenamefont {Lampen-Kelley}, \citenamefont {Knolle},
  \citenamefont {Balz}, \citenamefont {Aczel}, \citenamefont {Winn},
  \citenamefont {Liu}, \citenamefont {Pajerowski}, \citenamefont {Yan},
  \citenamefont {Bridges}, \citenamefont {Savici}, \citenamefont {Chakoumakos},
  \citenamefont {Lumsden}, \citenamefont {Tennant}, \citenamefont {Moessner},
  \citenamefont {Mandrus},\ and\ \citenamefont
  {Nagler}}]{banerjee_excitations_2018}%
  \BibitemOpen
  \bibfield  {author} {\bibinfo {author} {\bibfnamefont {A.}~\bibnamefont
  {Banerjee}}, \bibinfo {author} {\bibfnamefont {P.}~\bibnamefont
  {Lampen-Kelley}}, \bibinfo {author} {\bibfnamefont {J.}~\bibnamefont
  {Knolle}}, \bibinfo {author} {\bibfnamefont {C.}~\bibnamefont {Balz}},
  \bibinfo {author} {\bibfnamefont {A.~A.}\ \bibnamefont {Aczel}}, \bibinfo
  {author} {\bibfnamefont {B.}~\bibnamefont {Winn}}, \bibinfo {author}
  {\bibfnamefont {Y.}~\bibnamefont {Liu}}, \bibinfo {author} {\bibfnamefont
  {D.}~\bibnamefont {Pajerowski}}, \bibinfo {author} {\bibfnamefont
  {J.}~\bibnamefont {Yan}}, \bibinfo {author} {\bibfnamefont {C.~A.}\
  \bibnamefont {Bridges}}, \bibinfo {author} {\bibfnamefont {A.~T.}\
  \bibnamefont {Savici}}, \bibinfo {author} {\bibfnamefont {B.~C.}\
  \bibnamefont {Chakoumakos}}, \bibinfo {author} {\bibfnamefont {M.~D.}\
  \bibnamefont {Lumsden}}, \bibinfo {author} {\bibfnamefont {D.~A.}\
  \bibnamefont {Tennant}}, \bibinfo {author} {\bibfnamefont {R.}~\bibnamefont
  {Moessner}}, \bibinfo {author} {\bibfnamefont {D.~G.}\ \bibnamefont
  {Mandrus}},\ and\ \bibinfo {author} {\bibfnamefont {S.~E.}\ \bibnamefont
  {Nagler}},\ }\bibfield  {title} {\bibinfo {title} {Excitations in the
  field-induced quantum spin liquid state of {$\alpha-\text{RuCl}_3$}},\ }\href
  {https://doi.org/10.1038/s41535-018-0079-2} {\bibfield  {journal} {\bibinfo
  {journal} {npj Quantum Materials}\ }\textbf {\bibinfo {volume} {3}},\
  \bibinfo {pages} {8} (\bibinfo {year} {2018})}\BibitemShut {NoStop}%
\bibitem [{\citenamefont {Janssen}\ and\ \citenamefont
  {Vojta}(2019)}]{janssen_heisenbergkitaev_2019}%
  \BibitemOpen
  \bibfield  {author} {\bibinfo {author} {\bibfnamefont {L.}~\bibnamefont
  {Janssen}}\ and\ \bibinfo {author} {\bibfnamefont {M.}~\bibnamefont
  {Vojta}},\ }\bibfield  {title} {\bibinfo {title} {Heisenberg–{Kitaev}
  physics in magnetic fields},\ }\href
  {https://doi.org/10.1088/1361-648X/ab283e} {\bibfield  {journal} {\bibinfo
  {journal} {Journal of Physics: Condensed Matter}\ }\textbf {\bibinfo {volume}
  {31}},\ \bibinfo {pages} {423002} (\bibinfo {year} {2019})}\BibitemShut
  {NoStop}%
\bibitem [{\citenamefont {Zhu}\ \emph {et~al.}(2018)\citenamefont {Zhu},
  \citenamefont {Kimchi}, \citenamefont {Sheng},\ and\ \citenamefont
  {Fu}}]{zhu_robust_2018}%
  \BibitemOpen
  \bibfield  {author} {\bibinfo {author} {\bibfnamefont {Z.}~\bibnamefont
  {Zhu}}, \bibinfo {author} {\bibfnamefont {I.}~\bibnamefont {Kimchi}},
  \bibinfo {author} {\bibfnamefont {D.~N.}\ \bibnamefont {Sheng}},\ and\
  \bibinfo {author} {\bibfnamefont {L.}~\bibnamefont {Fu}},\ }\bibfield
  {title} {\bibinfo {title} {Robust non-{Abelian} spin liquid and a possible
  intermediate phase in the antiferromagnetic {Kitaev} model with magnetic
  field},\ }\href {https://doi.org/10.1103/PhysRevB.97.241110} {\bibfield
  {journal} {\bibinfo  {journal} {Phys. Rev. B}\ }\textbf {\bibinfo {volume}
  {97}},\ \bibinfo {pages} {241110} (\bibinfo {year} {2018})}\BibitemShut
  {NoStop}%
\bibitem [{\citenamefont {Gohlke}\ \emph {et~al.}(2018)\citenamefont {Gohlke},
  \citenamefont {Moessner},\ and\ \citenamefont
  {Pollmann}}]{gohlke_dynamical_2018}%
  \BibitemOpen
  \bibfield  {author} {\bibinfo {author} {\bibfnamefont {M.}~\bibnamefont
  {Gohlke}}, \bibinfo {author} {\bibfnamefont {R.}~\bibnamefont {Moessner}},\
  and\ \bibinfo {author} {\bibfnamefont {F.}~\bibnamefont {Pollmann}},\
  }\bibfield  {title} {\bibinfo {title} {Dynamical and topological properties
  of the {Kitaev} model in a [111] magnetic field},\ }\href
  {https://doi.org/10.1103/PhysRevB.98.014418} {\bibfield  {journal} {\bibinfo
  {journal} {Phys. Rev. B}\ }\textbf {\bibinfo {volume} {98}},\ \bibinfo
  {pages} {014418} (\bibinfo {year} {2018})}\BibitemShut {NoStop}%
\bibitem [{\citenamefont {Hickey}\ and\ \citenamefont
  {Trebst}(2019)}]{hickey_emergence_2019}%
  \BibitemOpen
  \bibfield  {author} {\bibinfo {author} {\bibfnamefont {C.}~\bibnamefont
  {Hickey}}\ and\ \bibinfo {author} {\bibfnamefont {S.}~\bibnamefont
  {Trebst}},\ }\bibfield  {title} {\bibinfo {title} {Emergence of a
  field-driven {U}(1) spin liquid in the {Kitaev} honeycomb model},\ }\href
  {https://doi.org/10.1038/s41467-019-08459-9} {\bibfield  {journal} {\bibinfo
  {journal} {Nature Communications}\ }\textbf {\bibinfo {volume} {10}},\
  \bibinfo {pages} {530} (\bibinfo {year} {2019})}\BibitemShut {NoStop}%
\bibitem [{\citenamefont {Zhang}\ \emph {et~al.}(2022)\citenamefont {Zhang},
  \citenamefont {Halász},\ and\ \citenamefont {Batista}}]{zhang_theory_2022}%
  \BibitemOpen
  \bibfield  {author} {\bibinfo {author} {\bibfnamefont {S.-S.}\ \bibnamefont
  {Zhang}}, \bibinfo {author} {\bibfnamefont {G.~B.}\ \bibnamefont {Halász}},\
  and\ \bibinfo {author} {\bibfnamefont {C.~D.}\ \bibnamefont {Batista}},\
  }\bibfield  {title} {\bibinfo {title} {Theory of the {Kitaev} model in a
  [111] magnetic field},\ }\href {https://doi.org/10.1038/s41467-022-28014-3}
  {\bibfield  {journal} {\bibinfo  {journal} {Nature Communications}\ }\textbf
  {\bibinfo {volume} {13}},\ \bibinfo {pages} {399} (\bibinfo {year}
  {2022})}\BibitemShut {NoStop}%
\bibitem [{\citenamefont {Ishikawa}\ \emph {et~al.}(2022)\citenamefont
  {Ishikawa}, \citenamefont {Kurihara}, \citenamefont {Yajima}, \citenamefont
  {Nishio-Hamane}, \citenamefont {Shimizu}, \citenamefont {Sakakibara},
  \citenamefont {Matsuo},\ and\ \citenamefont {Kindo}}]{ishikawa_SmI3_2022}%
  \BibitemOpen
  \bibfield  {author} {\bibinfo {author} {\bibfnamefont {H.}~\bibnamefont
  {Ishikawa}}, \bibinfo {author} {\bibfnamefont {R.}~\bibnamefont {Kurihara}},
  \bibinfo {author} {\bibfnamefont {T.}~\bibnamefont {Yajima}}, \bibinfo
  {author} {\bibfnamefont {D.}~\bibnamefont {Nishio-Hamane}}, \bibinfo {author}
  {\bibfnamefont {Y.}~\bibnamefont {Shimizu}}, \bibinfo {author} {\bibfnamefont
  {T.}~\bibnamefont {Sakakibara}}, \bibinfo {author} {\bibfnamefont
  {A.}~\bibnamefont {Matsuo}},\ and\ \bibinfo {author} {\bibfnamefont
  {K.}~\bibnamefont {Kindo}},\ }\bibfield  {title} {\bibinfo {title}
  {{$\text{SmI}_3$}:{$4f^5$} honeycomb magnet with spin-orbital entangled
  {$\Gamma_7$} {Kramers} doublet},\ }\href
  {https://doi.org/10.1103/PhysRevMaterials.6.064405} {\bibfield  {journal}
  {\bibinfo  {journal} {Phys. Rev. Mater.}\ }\textbf {\bibinfo {volume} {6}},\
  \bibinfo {pages} {064405} (\bibinfo {year} {2022})}\BibitemShut {NoStop}%
\bibitem [{\citenamefont {Georges}\ \emph {et~al.}(2013)\citenamefont
  {Georges}, \citenamefont {Medici},\ and\ \citenamefont
  {Mravlje}}]{georges_strong_2013}%
  \BibitemOpen
  \bibfield  {author} {\bibinfo {author} {\bibfnamefont {A.}~\bibnamefont
  {Georges}}, \bibinfo {author} {\bibfnamefont {L.~D.}\ \bibnamefont
  {Medici}},\ and\ \bibinfo {author} {\bibfnamefont {J.}~\bibnamefont
  {Mravlje}},\ }\bibfield  {title} {\bibinfo {title} {Strong {Correlations}
  from {Hund}’s {Coupling}},\ }\href
  {https://doi.org/10.1146/annurev-conmatphys-020911-125045} {\bibfield
  {journal} {\bibinfo  {journal} {Annual Review of Condensed Matter Physics}\
  }\textbf {\bibinfo {volume} {4}},\ \bibinfo {pages} {137} (\bibinfo {year}
  {2013})}\BibitemShut {NoStop}%
\bibitem [{\citenamefont {Anisimov}\ \emph {et~al.}(1997)\citenamefont
  {Anisimov}, \citenamefont {Aryasetiawan},\ and\ \citenamefont
  {Lichtenstein}}]{anisimov_first-principles_1997}%
  \BibitemOpen
  \bibfield  {author} {\bibinfo {author} {\bibfnamefont {V.~I.}\ \bibnamefont
  {Anisimov}}, \bibinfo {author} {\bibfnamefont {F.}~\bibnamefont
  {Aryasetiawan}},\ and\ \bibinfo {author} {\bibfnamefont {A.~I.}\ \bibnamefont
  {Lichtenstein}},\ }\bibfield  {title} {\bibinfo {title} {First-principles
  calculations of the electronic structure and spectra of strongly correlated
  systems: the {LDA}+ {U} method},\ }\href
  {https://doi.org/10.1088/0953-8984/9/4/002} {\bibfield  {journal} {\bibinfo
  {journal} {Journal of Physics: Condensed Matter}\ }\textbf {\bibinfo {volume}
  {9}},\ \bibinfo {pages} {767} (\bibinfo {year} {1997})}\BibitemShut {NoStop}%
\bibitem [{\citenamefont {Stevens}(1952)}]{stevens_matrix_1952}%
  \BibitemOpen
  \bibfield  {author} {\bibinfo {author} {\bibfnamefont {K.~W.~H.}\
  \bibnamefont {Stevens}},\ }\bibfield  {title} {\bibinfo {title} {Matrix
  {Elements} and {Operator} {Equivalents} {Connected} with the {Magnetic}
  {Properties} of {Rare} {Earth} {Ions}},\ }\href
  {https://doi.org/10.1088/0370-1298/65/3/308} {\bibfield  {journal} {\bibinfo
  {journal} {Proceedings of the Physical Society. Section A}\ }\textbf
  {\bibinfo {volume} {65}},\ \bibinfo {pages} {209} (\bibinfo {year}
  {1952})}\BibitemShut {NoStop}%
\bibitem [{\citenamefont {Lea}\ \emph {et~al.}(1962)\citenamefont {Lea},
  \citenamefont {Leask},\ and\ \citenamefont {Wolf}}]{lea_raising_1962}%
  \BibitemOpen
  \bibfield  {author} {\bibinfo {author} {\bibfnamefont {K.~R.}\ \bibnamefont
  {Lea}}, \bibinfo {author} {\bibfnamefont {M.~J.~M.}\ \bibnamefont {Leask}},\
  and\ \bibinfo {author} {\bibfnamefont {W.~P.}\ \bibnamefont {Wolf}},\
  }\bibfield  {title} {\bibinfo {title} {The raising of angular momentum
  degeneracy of f-{Electron} terms by cubic crystal fields},\ }\href
  {https://doi.org/https://doi.org/10.1016/0022-3697(62)90192-0} {\bibfield
  {journal} {\bibinfo  {journal} {Journal of Physics and Chemistry of Solids}\
  }\textbf {\bibinfo {volume} {23}},\ \bibinfo {pages} {1381} (\bibinfo {year}
  {1962})}\BibitemShut {NoStop}%
\bibitem [{\citenamefont {Takegahara}\ \emph {et~al.}(1980)\citenamefont
  {Takegahara}, \citenamefont {Aoki},\ and\ \citenamefont
  {Yanase}}]{takegahara_slater-koster_1980}%
  \BibitemOpen
  \bibfield  {author} {\bibinfo {author} {\bibfnamefont {K.}~\bibnamefont
  {Takegahara}}, \bibinfo {author} {\bibfnamefont {Y.}~\bibnamefont {Aoki}},\
  and\ \bibinfo {author} {\bibfnamefont {A.}~\bibnamefont {Yanase}},\
  }\bibfield  {title} {\bibinfo {title} {Slater-{Koster} tables for $f$
  electrons},\ }\href {https://doi.org/10.1088/0022-3719/13/4/016} {\bibfield
  {journal} {\bibinfo  {journal} {Journal of Physics C: Solid State Physics}\
  }\textbf {\bibinfo {volume} {13}},\ \bibinfo {pages} {583} (\bibinfo {year}
  {1980})}\BibitemShut {NoStop}%
\bibitem [{\citenamefont {Cox}\ \emph {et~al.}(1981)\citenamefont {Cox},
  \citenamefont {Lang},\ and\ \citenamefont {Baer}}]{cox_study_1981}%
  \BibitemOpen
  \bibfield  {author} {\bibinfo {author} {\bibfnamefont {P.~A.}\ \bibnamefont
  {Cox}}, \bibinfo {author} {\bibfnamefont {J.~K.}\ \bibnamefont {Lang}},\ and\
  \bibinfo {author} {\bibfnamefont {Y.}~\bibnamefont {Baer}},\ }\bibfield
  {title} {\bibinfo {title} {Study of the $4f$ and valence band density of
  states in rare-earth metals. {I}. {Theory} of the $4f$ states},\ }\href
  {https://doi.org/10.1088/0305-4608/11/1/014} {\bibfield  {journal} {\bibinfo
  {journal} {Journal of Physics F: Metal Physics}\ }\textbf {\bibinfo {volume}
  {11}},\ \bibinfo {pages} {113} (\bibinfo {year} {1981})}\BibitemShut
  {NoStop}%
\bibitem [{\citenamefont {Wybourne}\ and\ \citenamefont
  {Meggers}(1965)}]{wybourne_spectroscopic_1965}%
  \BibitemOpen
  \bibfield  {author} {\bibinfo {author} {\bibfnamefont {B.~G.}\ \bibnamefont
  {Wybourne}}\ and\ \bibinfo {author} {\bibfnamefont {W.~F.}\ \bibnamefont
  {Meggers}},\ }\bibfield  {title} {\bibinfo {title} {Spectroscopic
  {Properties} of {Rare} {Earths}},\ }\href {https://doi.org/10.1063/1.3047727}
  {\bibfield  {journal} {\bibinfo  {journal} {Physics Today}\ }\textbf
  {\bibinfo {volume} {18}},\ \bibinfo {pages} {70} (\bibinfo {year}
  {1965})}\BibitemShut {NoStop}%
\bibitem [{\citenamefont {Rusnačko}\ and\ \citenamefont
  {Chaloupka}(2019)}]{rusnaifmmode_checkcelse_cfiko_kitaev-like_2019}%
  \BibitemOpen
  \bibfield  {author} {\bibinfo {author} {\bibfnamefont {D.~G.}\ \bibnamefont
  {Rusnačko}, \bibfnamefont {Juraj}}\ and\ \bibinfo {author} {\bibfnamefont
  {J.}~\bibnamefont {Chaloupka}},\ }\bibfield  {title} {\bibinfo {title}
  {Kitaev-like honeycomb magnets: {Global} phase behavior and emergent
  effective models},\ }\href {https://doi.org/10.1103/PhysRevB.99.064425}
  {\bibfield  {journal} {\bibinfo  {journal} {Phys. Rev. B}\ }\textbf {\bibinfo
  {volume} {99}},\ \bibinfo {pages} {064425} (\bibinfo {year}
  {2019})}\BibitemShut {NoStop}%
\bibitem [{\citenamefont {Muniz}\ \emph {et~al.}(2014)\citenamefont {Muniz},
  \citenamefont {Kato},\ and\ \citenamefont
  {Batista}}]{GeneralizedSpinwave_Muniz_2014}%
  \BibitemOpen
  \bibfield  {author} {\bibinfo {author} {\bibfnamefont {R.~A.}\ \bibnamefont
  {Muniz}}, \bibinfo {author} {\bibfnamefont {Y.}~\bibnamefont {Kato}},\ and\
  \bibinfo {author} {\bibfnamefont {C.~D.}\ \bibnamefont {Batista}},\
  }\bibfield  {title} {\bibinfo {title} {Generalized spin-wave theory:
  {{Application}} to the bilinear-biquadratic model},\ }\href
  {https://doi.org/10.1093/ptep/ptu109} {\bibfield  {journal} {\bibinfo
  {journal} {Progress of Theoretical and Experimental Physics}\ }\textbf
  {\bibinfo {volume} {2014}},\ \bibinfo {pages} {83I01} (\bibinfo {year}
  {2014})}\BibitemShut {NoStop}%
\bibitem [{\citenamefont {Dong}\ \emph {et~al.}(2018)\citenamefont {Dong},
  \citenamefont {Wang},\ and\ \citenamefont {Li}}]{SUSpinwave_Dong_2018}%
  \BibitemOpen
  \bibfield  {author} {\bibinfo {author} {\bibfnamefont {Z.-Y.}\ \bibnamefont
  {Dong}}, \bibinfo {author} {\bibfnamefont {W.}~\bibnamefont {Wang}},\ and\
  \bibinfo {author} {\bibfnamefont {J.-X.}\ \bibnamefont {Li}},\ }\bibfield
  {title} {\bibinfo {title} {{{SU}}{$(N)$} spin-wave theory: {{Application}} to
  spin-orbital {{Mott}} insulators},\ }\href
  {https://doi.org/10.1103/PhysRevB.97.205106} {\bibfield  {journal} {\bibinfo
  {journal} {Physical Review B}\ }\textbf {\bibinfo {volume} {97}},\ \bibinfo
  {pages} {205106} (\bibinfo {year} {2018})}\BibitemShut {NoStop}%
\bibitem [{\citenamefont {Jang}\ and\ \citenamefont
  {Motome}(2024{\natexlab{b}})}]{jang_exploring_2024}%
  \BibitemOpen
  \bibfield  {author} {\bibinfo {author} {\bibfnamefont {S.-H.}\ \bibnamefont
  {Jang}}\ and\ \bibinfo {author} {\bibfnamefont {Y.}~\bibnamefont {Motome}},\
  }\bibfield  {title} {\bibinfo {title} {Exploring rare-earth {Kitaev} magnets
  by massive-scale computational analysis},\ }\href
  {https://doi.org/10.1038/s43246-024-00634-w} {\bibfield  {journal} {\bibinfo
  {journal} {Communications Materials}\ }\textbf {\bibinfo {volume} {5}},\
  \bibinfo {pages} {192} (\bibinfo {year} {2024}{\natexlab{b}})}\BibitemShut
  {NoStop}%
\end{thebibliography}%

\end{document}